\documentclass[conference]{IEEEtran}
\IEEEoverridecommandlockouts

\usepackage{cite}
\usepackage{amsmath,amssymb,amsfonts}
\usepackage{algorithm, algorithmic}
\usepackage{graphicx}
\usepackage{textcomp}
\usepackage{xcolor}
\usepackage{caption}
\usepackage{booktabs}

\def\BibTeX{{\rm B\kern-.05em{\sc i\kern-.025em b}\kern-.08em
    T\kern-.1667em\lower.7ex\hbox{E}\kern-.125emX}}

\usepackage{amsthm}
\newtheoremstyle{bfnote}%
{}{}
{\itshape}{}
{\bfseries}{.}
{ }{\thmname{#1}\thmnumber{ #2}\thmnote{ (#3)}}
\theoremstyle{bfnote}

\newtheorem{theorem}{Theorem}

\newtheorem{remark}{Remark}
\newtheorem{definition}{Definition}

\newcommand{\C}{\mathcal{C}} 
\newcommand{\Predicate}{\mathbf{P}} 
\newcommand{\Variable}{\mathbf{X}} 
\newcommand{\Constant}{\mathbf{C}} 
\newcommand{\AttConst}{\mathbf{C}_\mathbf{A}} 

\newcommand{\QSent}{\mathbf{Q}} 
\newcommand{\Language}{\mathcal{L}} 
\newcommand{\ProbMeas}{\mathcal{P}_{I}} 

\begin{document}

\title{Goal-Oriented Semantic Communication for Logical Decision Making}

\author{\IEEEauthorblockN{Ahmet Faruk Saz}
\IEEEauthorblockA{\textit{Dept. of Electrical and Computer Engineering} \\
\textit{Georgia Institute of Technology}\\
Atlanta, GA, USA \\
asaz3@gatech.edu}
\and
\IEEEauthorblockN{Faramarz Fekri}
\IEEEauthorblockA{\textit{Dept. of Electrical and Computer Engineering} \\
\textit{Georgia Institute of Technology}\\
Atlanta, GA, USA \\
faramarz.fekri@ece.gatech.edu}
}

\maketitle

\begin{abstract}
This paper develops a principled foundation for goal-oriented semantic communication for logical decision-making. Consider a setting where autonomous agents engage in collaborative perception. In such settings, the volume of sensory data and limited bandwidth often make transmission of raw observations infeasible, requiring intelligent selection of task-relevant information. Because these scenarios are safety-critical, the selection and decision processes must also be transparent and verifiable. To address this, we propose an explainable semantic communication framework grounded in a First-Order Logic (FOL) hierarchical representation of the world. We define semantic information, entropy, relative entropy, change in entropy, and mutual information by assigning an inductive logical probability measure over semantic structures in the language. Based on these definitions, we formulate a goal-oriented semantic communication objective through semantic rate-distortion theory and, equivalently, through the semantic information bottleneck principle. In this framework, task rules are represented as goal-oriented states, defined as a layer over the world states to capture decision-relevant abstractions. The resulting principle selects evidence that is most informative about these states, aiming to transmit only those FOL clauses most critical for decision-making while preserving logical verifiability. We demonstrate the effectiveness of the approach in a deduction-based safe path-following task within an FOL-based urban environment simulator with multiple dynamic agents.

\end{abstract}


\section{Introduction}
Semantic and goal-oriented communication have emerged as foundational paradigms for AI-native wireless systems, where the objective shifts from faithful reconstruction of raw data to transmission of meaning sufficient for correct downstream decision-making. In safety-critical domains such as autonomous driving, communicating all perceptual information is infeasible under bandwidth and latency constraints; agents must exchange only those observations that materially affect the receiver's actions. A principled framework must therefore answer two questions: \emph{what} to transmit, and \emph{why} that selection is correct---the latter being essential for verifiability in safety-critical settings.

Existing approaches fall into two categories. Black-box deep learning systems~\cite{Xie2020DeepLE, Yang2023SwinJSCCTS} achieve impressive bandwidth reductions but rely on opaque encodings. Mathematically grounded approaches~\cite{Niu2024AMT, Shao2022ATO, Niu2025RateDistortionPerceptionTI}---including synonymous mapping-based information theory, language-theoretic distortion-cost formulations, and rate-distortion-perception-semantics tradeoff analyses---offer principled guarantees but have not been unified into a framework that is simultaneously explainable, grounded in a formal theory of meaning, and tailored to logical decision-making. 

As part of latter efforts, in our former work, we proposed a FOL-based semantic information and communication framework for the representation of the state of the world. FOL provides a universal way of representing multi-modal information, as evidenced in knowledge graphs, OWL/RDF/XML databases, semantic web ontologies, and semantic sensor networks. Furthermore, FOL allows for logical reasoning over the data \cite{Krygin2024PerformingFQ}. We presented formulations for communicating the state of the world to a receiver for deduction \cite{Saz2025AnalysisOS}, and provided a mathematical analysis of the underlying semantic information and communication frameworks \cite{Saz2024OnTT}.

In this paper, we further our former work by presenting a goal-oriented semantic communication framework grounded in FOL and inductive logical probability. We introduce definitions for semantic content-information, entropy, relative entropy, change in entropy, and mutual information. We then formulate task-agnostic semantic communication as a semantic rate-distortion problem, then extend it to the goal-oriented setting via \emph{goal-oriented states}---logical equivalence classes induced by task rules. The objective becomes maximizing the change in semantic entropy of goal states due to transmitted evidence, equivalently expressed as a semantic information bottleneck. We propose a polynomial time lexicographical FOL expression sorting algorithm for the optimization of the goal-oriented semantic communication objective.

We demonstrate effectiveness in a deduction-based safe path-following task within the LogiCity simulator \cite{Li2024LogiCityAN}, where agents make safety-critical navigation decisions under communication constraints. Across multiple rule sets, architectures, and density configurations, semantic selection consistently outperforms random selection, with the advantage most pronounced under tight communication budgets.

\section{Background on FOL, Inductive Probabilities, and Semantic Information}

\begin{figure}[h]
    \centering
    \includegraphics[width=\linewidth]{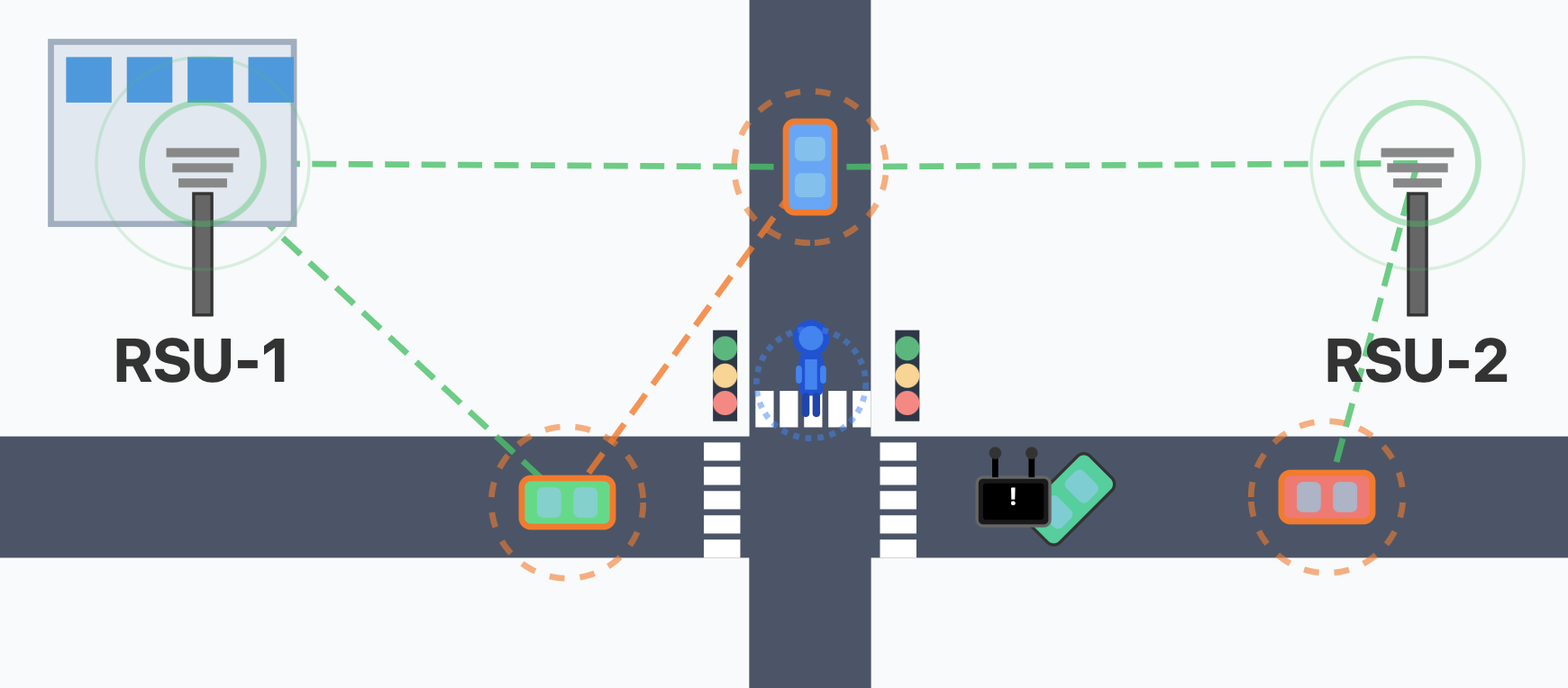}
    \caption{Collaborative Perception in Autonomous Driving through Goal-Oriented Semantic Communication}
    \label{fig:collusion}
\end{figure}

We consider a setting in which multiple vehicles, each with partial observations of the environment, transmit their observations to a receiver (e.g., a roadside unit or edge server) for cooperative perception, as depicted in Fig.~\ref{fig:collusion}. The receiver then transmits back to each agent the information most relevant to its task. Our framework is grounded in a First-Order Logic (FOL) representation: a logic encoder $f_\zeta$ maps each vehicle's local perceptual state $V$ to a set of FOL clauses $e = f_\zeta(V)$, and a lossy semantic encoder $g_\theta(\hat{e} \mid e)$ selects a subset $\hat{e} \subseteq e$ for transmission in accordance with communication constraints.

\begin{remark}
Alternatively, it is possible to train a joint encoder $h_\theta(\hat{e} \mid V)$ that maps directly from perceptual observations to semantic encodings (i.e., logic clauses). 
\end{remark}

Following \cite{c4, c10, nelte}, let $\Language$ be a first-order language whose vocabulary consists of monadic and dyadic predicate symbols $\Predicate=\{P_i\}_{i=1}^k$, a countably infinite set of variables $\Variable$, constants $\Constant$, logical connectives $\{\neg,\land,\lor\}$, and quantifiers $\{\forall,\exists\}$. For example, the predicate $\textit{LeftOf}(x_1,x_2)$ evaluates to true if vehicle $x_1$ is to the left of vehicle $x_2$ and false otherwise. Through language $\Language$, we can express logical rules, e.g.,
\begin{equation}
\label{equ:rule1}
\begin{aligned}
\forall x_1 \;\big( \exists x_2 \;
 &InIntersection(x_1, x_2)\land IsPedestrian(x_2) \\
 &\rightarrow Stop(x_2, x_1)
\big),
\end{aligned}
\end{equation}
which asserts: “If there exists a pedestrian $x_2$ who is in the intersection, then all vehicles $x_1$ shall stop for pedestrian $x_2$.”.

\subsection{Q-sentences and Attributive Constituents}

To describe relational structure at the semantic level, we introduce \emph{Q-sentences}, each encoding a complete pattern of dyadic relations between two individuals.

\begin{definition}[Q-sentence]
For a configuration $i$, let $\delta_{i,P},\delta'_{i,P}\in\{\pm1\}$ for all $P\in\Predicate$\footnote{For monadic predicates, each predicate $P$ appears in the formulation of Q-sentence as $\delta_{i,P}P(x_1)$ by itself.}. For $x_1,x_2\in\Variable$,
\[
Q_i(x_1,x_2)
=\bigwedge_{P\in\Predicate}
\left(\delta_{i,P}P(x_1,x_2)\,\land\,\delta'_{i,P}P(x_2,x_1)\right),
\]
where $\delta_{i,P}=+1$ denotes $P(x_1,x_2)$ and $\delta_{i,P}=-1$ denotes $\neg P(x_1,x_2)$; similarly for $\delta'_{i,P}$. Each configuration $i$ therefore specifies one complete enumeration of predicate occurrences.
\end{definition}

There are $4^{|\Predicate_{dyadic}|}2^{|\Predicate_{monadic}|}$ Q-sentences, exactly one of which is satisfied under any valuation $v[x_1\mapsto e_1,x_2\mapsto e_2]$ (or $v[x_1\mapsto e_1]$). This observation motivates \emph{attributive constituents}, which specify the full set of Q-sentences realized by a single entity, thereby characterizing \emph{what kind of individual}  that entity is with respect to the relations it possesses or lacks.

\begin{definition}[Attributive Constituent]
Let $\QSent$ denote all Q-sentences. For $\delta_{j,Q}\in\{\pm1\}$,
\[
C_j(x_1)
=\bigwedge_{Q\in\QSent}
\delta_{j,Q}\big((\exists x_2)\,Q(x_1,x_2)\big),
\]
where $\delta_{j,Q}=+1$ asserts $(\exists x_2)Q(x_1,x_2)$ and $\delta_{j,Q}=-1$ asserts its negation.
\end{definition}

Thus, $C_j(x_1)$ encodes the exact set of relational patterns that $x_1$ satisfies relative to all others (through dyadic predicates) and properties $x_1$ possess (through monadic predicates). Distinct individuals may share the same attributive constituent; the total number of such constituents is at most $2^{4^{|\Predicate_{dyadic}|}2^{|\Predicate_{monadic}|}}$.

\subsection{Constituents and the State of the World}

To represent the entire structure of the entire universe in terms of the different kinds of individuals it contains, we combine attributive constituents into \emph{constituents}.

\begin{definition}[Constituent]
Let $\AttConst$ be the set of all attributive constituents. For $\delta_{k,C_A}\in\{\pm1\}$,
\[
C^{w,k}
=\bigwedge_{C_A\in\AttConst}
\delta_{k,C_A}\big((\exists x_1)C_A(x_1)\big),
\]
where the width $w$ is the number of conjuncts with $\delta_{k,C_A}=+1$. Here $\delta_{k,C_A}=+1$ asserts that some $x_1$ satisfies $C_A(x_1)$; $\delta_{k,C_A}=-1$ asserts that none do.
\end{definition}

There are at most $2^{2^{4^{|\Predicate_{dyadic}|}2^{|\Predicate_{monadic}|}}}$ constituents, exactly one of which is satisfied by the true (but unknown) world model. Let $\C$ denote all constituents. For the true world, exactly one constituent $C^*\in\C$ is satisfied, and this constituent uniquely represents the \emph{true state of the world}. In practice, as true constituent $C^*$ is unknown, $C^*$ is approximated by the constituent that is best supported by available evidence $e$.

Every closed generalization in $\Language$ admits an equivalent disjunctive normal form over constituents, establishing them as the semantic basis of the language.

\begin{theorem}[Distributive Normal Forms in FOL \cite{nelte,c9,c4}]\label{dnfinfol}
Let $\Language$ have finite dyadic predicates $\Predicate$, and let $\C$ denote all constituents. For any closed general sentence $\varphi$,
\[
\varphi\equiv
\bigvee_{\substack{C^{w,k}\in\C\\ C^{w,k}\models\varphi}}
C^{w,k}.
\]
\end{theorem}

\subsection{Inductive Probabilities and Content Information}

We place an inductive logical probability measure on the set of constituents $\C$. The prior over $\C$ represents a rational agent’s initial beliefs about truth of world states, and this prior is updated inductively as new evidence arrives, yielding posterior beliefs about the true state of the world.

\begin{remark}
    To seperate them from statistical probabilities, notation $c(\cdot)$ is used to denote inductive probabilities.
\end{remark}

Next, we introduce the \emph{degree of confirmation}, $c(\varphi \mid e)$, which quantifies the support that evidence $e$ provides for a hypothesis $\varphi$ and is a conditional logical probability.

\begin{definition}[Degree of Confirmation \cite{c2}]\label{degofconf}
For any FOL expression $\varphi$ and evidence $e$, the \emph{conditional inductive logical probability} (degree of confirmation) of $\varphi$ given $e$ is
\[
c(\varphi \mid e)
= \frac{c(\varphi \land e)}{c(e)}.
\]
\end{definition}

Thus the Bayesian posterior over constituents is
\[
\ProbMeas
=\{\,c(C^{w,k},e): C^{w,k}\in\C\,\},
\qquad
\sum_{C^{w,k}\in\C}c(C^{w,k},e)=1.
\]

Different inductive systems—such as Carnap’s $\lambda$-continuum~\cite{c1}, Hintikka's two-dimensional framework~\cite{c10}, Hilpinen's relational framework~\cite{c4}—assign probabilities to constituents in fundamentally different ways. The choice of inductive system therefore has direct consequences for the resulting posterior distribution over world constituents and, consequently, for all derived semantic inferences.

In this work, we adopt an inductive system, in which the likelihood is deterministic:
\begin{equation}\label{indsys1}
c(e \mid C^{w,k}) :=L(e \mid C^{w,k}) = \mathbf{1}\{C^{w,k} \models e\}.
\end{equation}
That is, a constituent either satisfies the evidence or it does not. Applying Bayes' rule with prior $\c(C^{w,k})$ yields the posterior degree of confirmation
\begin{equation}\label{indsys2}
c(C^{w,k} \mid e) = \frac{\mathbf{1}\{C^{w,k} \models e\}\,c(C^{w,k})}{\sum_{C^{w',k'} \in \C} \mathbf{1}\{C^{w',k'} \models e\}\,c(C^{w',k'})}.
\end{equation}
Under the assumption of symmetry, the prior is $c(C^{w,k}) = 1/|\C|$ for all $C^{w,k} \in \C$. Defining the compatible set $\C(e) := \{C^{w,k} \in \C : C^{w,k} \models e\}$, the posterior simplifies to
\begin{equation}\label{indsys3}
c(C^{w,k} \mid e) =
\begin{cases}
1/|\C(e)|, & C^{w,k} \models e,\\
0, & \text{otherwise},
\end{cases}
\qquad
c(e) = \frac{|\C(e)|}{|\C|}. 
\end{equation}
Every constituent compatible with $e$ receives equal posterior probability, and every incompatible constituent receives zero. This posterior over constituents induces the degree of confirmation for arbitrary sentences. Since each constituent is a complete world-description, it either satisfies a hypothesis $\varphi$ or does not. Applying Definition~\ref{degofconf},
\begin{equation}\label{indsys4}
c(\varphi \mid e) = \sum_{C^{w,k} \in \C} \mathbf{1}\{C^{w,k} \models \varphi\}\,c(C^{w,k} \mid e), 
\end{equation}
which under direct uniformity reduces to
\begin{equation}\label{indsys5}
c(\varphi \mid e) = \frac{|\C(e \wedge \varphi)|}{|\C(e)|} \qquad c(\varphi) = \frac{|\C(\varphi)|}{|\C|},
\end{equation}
where $\C(e \wedge \varphi) := \{C^{w,k} \in \C : C^{w,k} \models e \wedge \varphi\}$. Thus, all conditional and unconditional inductive probabilities reduce to ratios of constituent counts, and the posterior $\ProbMeas$ is fully determined by the logical compatibility structure of $\Language$.

Carnap, based on the inductive probability distribution over world states introduced the notion of \emph{content-information}, quantifying the \emph{information content of statements}.

\begin{definition}[Content Information \cite{c2}]\label{cont-info}
For any hypotheses $\varphi$, its content-information is
\[
\mathrm{cont}(\varphi)=1-c(\varphi).
\]
Given a fixed hypothesis $\varphi$ and evidence $e$, the conditional content information of evidence $e$ with respect to $\varphi$ is measured as:
\[
\mathrm{cont}(\varphi \mid e) = \mathrm{cont}(\varphi \land e) - \mathrm{cont}(e),
\]
\end{definition}

The $\mathrm{cont}$ metric admits two complementary readings. In its absolute form, $\mathrm{cont}(\varphi) = 1 - c(\varphi)$ quantifies how many possible world states a statement eliminates if true. Consider a toy propositional world with two predicates $s$ and $r$, yielding four equiprobable states. The conjunction $\varphi_1 \equiv s \land r$ rules out three of the four states, giving $\mathrm{cont}(\varphi_1) = 0.75$, whereas $\varphi_2 \equiv s$ rules out only two, giving $\mathrm{cont}(\varphi_2) = 0.5$. Statements that are harder to satisfy---and hence less likely to be true---carry higher content, as their truth eliminates more alternatives. Put simply, the more unlikely a statement, the greater its information content. Conditioned on evidence $e$, $
\mathrm{cont}(\varphi \mid e)
= \mathrm{cont}(\varphi \land e)-\mathrm{cont}(e)
= c(e)-c(\varphi\land e),
$ measures the information content of $\varphi$ relative to $e$. If $\varphi$ tells us essentially what is already contained in the evidence—that is, if $\mathrm{cont}(\varphi\land e)\approx \mathrm{cont}(e)$—then $\varphi$ largely confirms or reinforces what is already known. Its truth is therefore only weakly surprising and, correspondingly, weakly informative. By contrast, if $\varphi$ conveys information not already contained in the evidence, thereby eliminating additional possible states—that is, if $\mathrm{cont}(\varphi\land e)\ll \mathrm{cont}(e)$—then its truth is more surprising and hence more informative. 

In the next section, we introduce different forms of semantic content information.

\section{Types of Semantic Content Information}

Logical probability---i.e., degree of confirmation---as defined in \eqref{degofconf} is inherently conditional, with its two arguments conventionally called a hypothesis and evidence.

\begin{remark}
    These labels are conventional: any FOL sentences or sets of sentences may occupy these roles, with degree of confirmation measuring how strongly the evidence supports the hypothesis. In logical probability, all available evidence is represented by a single sentence $e$, formed by conjoining individual observations. Carnap \cite{carnap_logical_1962} explicitly separates methodological considerations---such as measurement procedures and noisy observations---from logical probability. Evidence is therefore not a statistical random variable, but an FOL observation report. Consequently, in defining types of information below, we do not take an expectation or average over the evidence. Hypotheses, initially uncertain, correspond to the possible FOL descriptions of an unknown. Once the true hypothesis is identified, it becomes part of the evidence.
\end{remark}

We first consider uncertainty over a set of hypotheses $\{\varphi_i\}_i$. Under tautological evidence $e\equiv\top$, its semantic content entropy is
\begin{equation}
\mathrm{H_s}(\mathbf{\varphi})
=\mathbb{E}_{\varphi}[\mathrm{cont}(\varphi)]
=\sum_{i} c(\varphi_i)\,\mathrm{cont}(\varphi_i),
\end{equation}
which measures average a priori uncertainty. Statements believed with high confidence, whether true or false, contribute little information, whereas statements that are maximally uncertain contribute the most information.

Given evidence $e$, the relative semantic content entropy is
\begin{align}
\mathrm{H_s}(\mathbf{\varphi}\mid e)
&=\mathbb{E}_{\varphi}
   [\mathrm{cont}(\varphi\mid e)] \nonumber\\
&=\sum_i c(\varphi_i\mid e)\,
   \mathrm{cont}(\varphi_i\mid e),
\end{align}
measuring the uncertainty remaining in $\varphi$ after observing $e$. If $e$ strongly discriminates among the hypotheses---confirming a few and ruling out the rest---the posterior is concentrated and the conditional entropy is low. Conversely, if $e$ is uninformative, posterior remains diffuse and conditional entropy is high.

This, however, is not the only notion of relative semantic content entropy. Suppose we are given another set of hypotheses $\{\psi_j\}_j$ and wish to characterize how much uncertainty, on average, remains regarding the first set of hypotheses $\{\varphi_i\}_i$ relative to the second set $\{\psi_j\}_j$, under common evidence $e$. Then, relative semantic content entropy is defined as:
\begin{equation}\label{relsement}
\mathrm{H_s(\mathbf{\varphi}/\mathbf{\psi}\mid e)}
=
\mathrm{H_s(\mathbf{\varphi}\land\mathbf{\psi}\mid e)}
-
\mathrm{H_s(\mathbf{\psi}\mid e)}.
\end{equation}
Equivalently,
\begin{align}
\mathrm{H_s(\mathbf{\varphi}/\mathbf{\psi}\mid e)}
&=
\mathbb{E}_{\psi,\varphi}
[\mathrm{cont}(\mathbf{\varphi}\land\mathbf{\psi}\mid e)]
-
\mathbb{E}_{\psi}
[\mathrm{cont}(\mathbf{\psi}\mid e)]
\nonumber\\
&=
\mathbb{E}_{\psi,\varphi}
[\mathrm{cont}(\varphi\mid e\land\psi)]
\nonumber\\
&=
\mathbb{E}_{\psi}
[\mathrm{H_s}(\mathbf{\varphi}\mid e\land\psi)]
\nonumber\\
&=
\mathrm{H'_s}(\mathbf{\varphi}\mid e\land\psi),
\end{align}
where
\begin{align}
\mathrm{H'_s}(\mathbf{\varphi}\mid e\land\psi)
&=
\sum_j c(\psi_j\mid e)
\mathrm{H_s}(\mathbf{\varphi}\mid e\land\psi_j)
\nonumber\\
&=
\sum_{i,j} c(\psi_j\mid e)c(\varphi_i\mid e\land\psi_j)
\mathrm{cont}(\varphi_i\mid e\land\psi_j)
\nonumber\\
&=
\sum_{i,j} c(\varphi_i\land\psi_j\mid e)
\mathrm{cont}(\varphi_i\mid e\land\psi_j).
\end{align}
Thus, the intuitive interpretation of
$\mathrm{H_s (\mathbf{\varphi} / \mathbf{\psi}| e)}$ is the average uncertainty, or information, associated with the first set of hypotheses when we know both the evidence $e$ and the truth of a particular outcome from the hypothesis set $\{\psi_j\}_j$. As the actually true $\psi_j$ is itself unknown, this remaining uncertainty is averaged over all possible outcomes $\{\psi_j\}_j$.

For example, if $\mathbf{\varphi}$ describe whether a driver is careful, average, or reckless, and $\mathbf{\psi}$ whether they have had no crashes, $1$--$3$, or more than $3$; \eqref{relsement} characterizes the remaining uncertainty about driver classification given the crash category and background statistics, averaged over all crash categories.

These two notions of relative semantic content entropy yield two related notions of semantic mutual content information. The first, which we call the \emph{change in semantic content entropy}, measures the effect of evidence on a hypothesis set:
\begin{align}\label{changeinsement}
\mathrm{\gamma_s(\mathbf{\varphi}\mid e)}
&=
\mathbb{E}_{\varphi}
[\mathrm{cont}(\varphi)-\mathrm{cont}(\varphi\mid e)]
\nonumber\\
&=
\sum_i c(\varphi_i)\mathrm{cont}(\varphi_i)
-
\sum_i c(\varphi_i\mid e)\mathrm{cont}(\varphi_i\mid e)
\nonumber\\
&=
\sum_i c(\varphi_i)\mathrm{cont}(\varphi_i)
-
\sum_i c(\varphi_i\mid e)
[c(e)-c(\varphi_i\land e)]
\nonumber\\
&=
\sum_i c(\varphi_i)\mathrm{cont}(\varphi_i)
-
\sum_i c(e)c(\varphi_i\mid e)
[1-c(\varphi_i\mid e)]
\nonumber\\
&=
\mathrm{H_s(\mathbf{\varphi})}
-
\mathrm{H_s(\mathbf{\varphi}\mid e)}.
\end{align}
Hence, $\gamma_s(\mathbf{\varphi}\mid e)$ quantifies the reduction in uncertainty about $\mathbf{\varphi}$ induced by evidence $e$.

The second notion measures the mutual semantic content between two hypothesis sets:
\begin{align}\label{semmutstand}
\mathrm{I_s(\mathbf{\varphi};\mathbf{\psi}\mid e)}
&=
\mathrm{H_s(\mathbf{\varphi}\mid e)}
+
\mathrm{H_s(\mathbf{\psi}\mid e)}
-
\mathrm{H_s(\mathbf{\varphi}\land\mathbf{\psi}\mid e)}
\nonumber\\
&=
\mathrm{H_s(\mathbf{\varphi}\mid e)}
-
\mathrm{H_s(\mathbf{\varphi}/\mathbf{\psi}\mid e)}
\nonumber\\
&=
\mathrm{H_s(\mathbf{\psi}\mid e)}
-
\mathrm{H_s(\mathbf{\psi}/\mathbf{\varphi}\mid e)}.
\end{align}
This has the same structural form as Shannon mutual information. Using the previous decomposition,
\begin{align}\label{semmutequiv}
\mathrm{I_s(\mathbf{\varphi};\mathbf{\psi}\mid e)}
&=
\mathrm{H_s(\mathbf{\varphi}\mid e)}
-
\mathrm{H'_s}(\mathbf{\varphi}\mid e\land\psi)
\nonumber\\
&=
\mathrm{H_s(\mathbf{\psi}\mid e)}
-
\mathrm{H'_s}(\mathbf{\psi}\mid e\land\varphi).
\end{align}
Equation \eqref{semmutequiv} therefore measures the expected reduction in uncertainty about $\mathbf{\varphi}$ if an outcome $\psi_j$ were observed and added to the existing evidence $e$, averaged over all possible $\psi_j$.

If the true outcome $\psi_j^*$ becomes known, this expected reduction becomes an ordinary change in semantic content entropy, but relative to prior evidence $e$ rather than tautological evidence. The relation to \eqref{changeinsement} can be made explicit as
$\mathrm{\gamma_s(\mathbf{\varphi}\mid e)}
=
\mathrm{H_s(\mathbf{\varphi})}
-
\mathrm{H_s(\mathbf{\varphi}\mid e)}
=
\mathrm{H_s(\mathbf{\varphi}\mid\top)}
-
\mathrm{H_s(\mathbf{\varphi}\mid e\land\top)}
=
\mathrm{I_s(\mathbf{\varphi};\top\mid e)}.
$
Accordingly, \eqref{changeinsement} considers new evidence $e$ starting from tautological evidence, whereas \eqref{semmutequiv} considers additional evidence $\psi_j$ on top of already available evidence $e$. When $\psi_j^*$ becomes known,
$\mathrm{\gamma'_s(\mathbf{\varphi}\mid\psi_j^*)}
=
\mathrm{I_s(\mathbf{\varphi};\psi_j^*\mid e)}
=
\mathrm{H_s(\mathbf{\varphi}\mid e)}
-
\mathrm{H_s(\mathbf{\varphi}\mid e\land\psi_j^*)}.
$ Finally, under tautological evidence, \eqref{semmutequiv} becomes
\begin{align}\label{shannonmutinsem}
\mathrm{I_s(\mathbf{\varphi};\mathbf{\psi})}
&=
\mathrm{H_s(\mathbf{\varphi})}
-
\mathrm{H'_s}(\mathbf{\varphi}\mid\psi)
\nonumber\\
&=
\mathbb{E}_{\varphi}[\mathrm{cont}(\varphi)]
-
\mathbb{E}_{\psi,\varphi}
[\mathrm{cont}(\varphi\mid\psi)]
\nonumber\\
&=
\sum_i c(\varphi_i)\mathrm{cont}(\varphi_i)
-
\sum_{i,j}c(\varphi_i\land\psi_j)
\mathrm{cont}(\varphi_i\mid\psi_j).
\end{align}
For two sets of hypotheses---for example, the possible outcomes of two experiments not yet conducted---\eqref{shannonmutinsem} quantifies \emph{a priori} how informative the outcome of one is about the other. Its form is identical to Shannon mutual information.

\section{Goal-Oriented Semantic Communication}
\label{sec:goal_oriented}

In our former work~\cite{Saz2024LossySC, Saz2025AnalysisOS}, we studied semantic communication schemes that transmit observations so as to most effectively reduce uncertainty over the \emph{state of the world}, represented by constituents in~$\C$. In that task-agnostic setting, the receiver's objective is to recover the evidence (i.e., observations) as faithfully as possible under rate constraints. This problem admits a semantic rate--distortion formulation, which we briefly recall below.

\begin{theorem}[Semantic Rate--Distortion Theorem]
\label{thm:semantic-rate-distortion}
Let $e$ denote the complete set of observations and let $\hat{e}$ be a subset of evidence $e$. Let $d_s(e, \hat{e})$ be a bounded semantic distortion function. For a semantic distortion constraint $D$, the minimum achievable semantic rate is
\begin{equation}
R_s(D)
=
\min_{\substack{\hat{e} \subseteq e:\\\\ d_s(e,\hat{e}) \le D}}
\mathrm{\gamma_s}(e\mid\hat{e}),
\end{equation}
The distortion $d_s(e,\hat{e})$ can be selected to count changes in the disjunctive normal form representation of $e$: information loss may add or remove constituents, and both effects incur semantic distortion.
\end{theorem}

\begin{remark}
A secondary compression layer may be applied to $\hat{e}$ (e.g., via an autoencoder or joint source--channel coding). The semantic rate--distortion characterisation extends without conceptual changes.
\end{remark}

\begin{remark}
    If the objective of the receiver is to determine the state of the world as accurately as possible, the mutual information term in Theorem \ref{thm:semantic-rate-distortion} is replaced with $\mathrm{\gamma_s}(\C\mid\hat{e})$ where $\C$ is the set of constituents.
\end{remark}

In many decision-making settings, however, identifying the exact state of the world is unnecessary. What ultimately matters is how evidence supports \emph{correct task execution}---such as deduction, prediction, induction, abduction, or counterfactual reasoning. Evidence is therefore valuable only insofar as it improves the receiver's ability to make decisions and take appropriate actions.

Let $L_{\mathcal{T}}$ denote a collection of logical rules encoding task priorities and constraints (e.g., traffic rules or safety requirements). These rules induce a logical equivalence relation (as per Theorem \ref{dnfinfol}) over constituents in~$\C$, grouping together all world states that lead to the same task outcome.

\begin{definition}[Goal-Oriented States]
For each subset of constituents $\C_m \subseteq \C$ induced by $L_{\mathcal{T}}$, define the corresponding goal-oriented state as
\[
\varphi_m \equiv \bigvee_{C^{w,k} \in \C_m} C^{w,k}.
\]
The collection $\Phi = \{\varphi_1, \dots, \varphi_M\}$ is called the \emph{goal-state space}. Each goal state $\varphi_m \in \Phi$ is associated with a corresponding action $a_m$ from an action set $\mathcal{A} = \{a_1, \dots, a_R\}$.
\end{definition}

Goal states need not be mutually exclusive---$\varphi_k \land \varphi_\ell \not\equiv \bot$ for some $k \neq \ell$---but they are jointly exhaustive:
\[
\bigvee_{m=1}^{M} \varphi_m \;\equiv\; \bigvee_{C^{w,k} \in \C} C^{w,k}.
\]

\begin{definition}[Goal-Oriented Inductive Probability]
Given evidence $e$, the inductive probability of a goal state $\varphi_m$ is
\[
c(\varphi_m \mid e) = \sum_{C^{w,k} \models \varphi_m} c(C^{w,k} \mid e).
\]
\end{definition}

In the context of autonomous driving, the task rules $L_{\mathcal{T}}$ take the form of expressions such as~(\ref{equ:rule1}), specifying the conditions under which each action is appropriate. The goal-oriented states $\Phi$ are then defined as the triggering conditions of these rules with respect to an ego vehicle $\mathrm{ego}$, i.e., 
\begin{equation}\label{hypo1}
    \varphi_1(\mathrm{ego}) = \exists x_2 \;InIntersection(\mathrm{ego}, x_2)\land IsPedestrian(x_2)
\end{equation}

The corresponding action is then $a_1 = Stop$ for $\varphi_1$. 

\begin{remark}
Although each goal state $\varphi_m$ admits a constituent decomposition via Theorem~\ref{dnfinfol}, i.e., $\varphi_m \equiv \bigvee_{C^{w,k} \models \varphi_m} C^{w,k}$, in practice the goal states are grounded with respect to a specific ego vehicle. This is because we are typically not interested in the information that evidence conveys, on average, about the truth of a hypothesis for every vehicle---as the universal quantification in~(\ref{equ:rule1}) would suggest---but rather in evaluating the truth of each hypothesis with respect to a particular ego vehicle of interest.
\end{remark}

Evidence is then \emph{task-informative} precisely when it concentrates posterior probability mass on a small subset of $\Phi$ or eliminates irrelevant goal states. This motivates the following semantic information bottleneck formulation for goal-oriented semantic communication.

\begin{theorem}[Goal-Oriented Semantic Communication Principle]
\label{thm:goal-oriented-semantic}
Let $\Phi = \{\varphi_1, \dots, \varphi_M\}$ be the goal-state space induced by $L_{\mathcal{T}}$. Given evidence $e$, the optimal transmitted evidence for goal-oriented semantic communication is any solution of
\begin{equation}
\label{eq:goal-ib}
\max_{g_\theta} \; \mathrm{\gamma_s}(\Phi \mid \hat{e}) \;-\; \beta\, \mathrm{I}(e; \hat{e}),
\end{equation}
where $\mathrm{I}(\cdot)$ is Shannon mutual information. 
\end{theorem}

Semantic information bottleneck requires transmitting as little as possible while retaining maximal semantic mutual content with the goal states. Since the goal-state space $\Phi$ is fixed, the semantic content entropy $\mathrm{H_s}(\Phi)$ is constant. Maximizing semantic mutual content $\mathrm{I_s}(\Phi; \hat{e})$ is therefore equivalent to minimizing the conditional semantic content entropy $\mathrm{H_s}(\Phi \mid \hat{e})$. A single goal state contributes $c(\varphi_m, e)\,\mathrm{cont}(\varphi_m \mid e)$ to this entropy, which is small when $\varphi_m$ is nearly certainly true ($c \approx 1$, $\mathrm{cont} \approx 0$) or nearly certainly false ($c \approx 0$, $\mathrm{cont} \approx 1$), and maximized at $c = \mathrm{cont} = 0.5$. The optimal evidence is therefore that which most sharply resolves the truth or falsity of each goal state.

\section{Goal-Oriented Semantic Communication Algorithm}
\label{sec:algorithm}

To optimize the goal-oriented semantic communication objective~(\ref{eq:goal-ib}), we enforce the rate constraint $\mathrm{I}(e;\hat{e})$ by fixing the number of FOL sentences that can be transmitted. Since the vocabulary of $\Language$ is known to both transmitter and receiver, each FOL-based perceptual observation admits a fixed-length encoding, so constraining the subset size to $|\hat{e}| = k$ directly bounds the transmission rate. Under the inductive system adopted in this paper (eqn. (\ref{indsys1})-(\ref{indsys5})), the objective~(\ref{eq:goal-ib}) then reduces to selecting the evidence subset $\hat{e} \subseteq e$ of size $k$ that minimizes the total weighted uncertainty over goal states $\Phi = \{\varphi_1, \dots, \varphi_M\}$:
\begin{equation}
\hat{e}^* = \arg\min_{\substack{\hat{e} \subseteq e \\ |\hat{e}| = k}} \; \mathcal{F}(\hat{e}), \quad
\mathcal{F}(\hat{e}) = \sum_{i=1}^{M} c(\hat{e}) \cdot c(\varphi_i \mid \hat{e}) \cdot \bigl(1 - c(\varphi_i \mid \hat{e})\bigr).
\label{eq:objective}
\end{equation}

Each summand is the Bernoulli variance of $\varphi_i$ under the transmitted evidence, weighted by $c(\hat{e})$. For hypothesis of kind \ref{hypo1}) and inductive system presented in eqn.~(\ref{indsys1})--(\ref{indsys5}), a goal state $\varphi_i$ is said to \emph{overlap} with the evidence $\hat{e}$ if at least one Q-sentence in $\hat{e}$ satisfies all predicate slots fixed by $\varphi_i$; in this case, the evidence directly witnesses the hypothesis, $c(\varphi_i \mid \hat{e}) = 1$, and the corresponding term vanishes ($\mathcal{F}_i = 0$). Otherwise, $\varphi_i$ is \emph{non-overlapping}, and its conditional probability takes the form $c(\varphi_i \mid \hat{e}) = (1 - u_i)/(1 - v)$. Substituting the closed-form expressions from the inductive system (eqn.~(\ref{indsys1})--(\ref{indsys5}))---where $v \triangleq 2^{-\alpha}$, $u_i \triangleq 2^{-\gamma_i}$, $\alpha \triangleq 2^{|\QSent|-K}$, and $\gamma_i \triangleq 2^{|\QSent|-K} - 2^{|\QSent|-K-H_i}$---each non-overlapping term simplifies to
$\mathcal{F}_i = \frac{(1 - u_i)(u_i - v)}{1 - v}.$
Here $K$ is the number of distinct Q-sentences in the evidence, and $H_i = 2^{T - Z_i}$ is the number of Q-sentences compatible with $\varphi_i$ (which fixes $Z_i$ predicate slots out of $T$ possible predicate slots).

For the language parameters in our application, the quantities $\alpha$ and $\gamma_i$ are integers with billions of digits, so $u_i$ and $v$ underflow to zero in floating-point arithmetic. However, in this regime $\mathcal{F}_i \approx u_i = 2^{-\gamma_i}$, and the sum is dominated by the smallest exponent: $\mathcal{F}(\hat{e}) \approx 2^{-\gamma_{\min}}$ where $\gamma_{\min} = \min_{i \in \mathcal{I}_{\not\sim}} \gamma_i$ and $\mathcal{I}_{\not\sim}$ indexes the non-overlapping goal states. Minimising $\mathcal{F}$ therefore reduces to maximising $\gamma_{\min}$.

Factoring $\gamma_i = 2^{|\QSent|-K-H_i}(2^{H_i} - 1)$ reveals two hierarchically ordered effects: (1)~\emph{evidence diversity}---smaller $K$ increases every $\gamma_i$ doubly exponentially, dominating all other factors; and (2)~\emph{hypothesis specificity}---for equal $K$, the bottleneck goal state with the smallest $H_i$ determines $\gamma_{\min}$, and maximising $H_{\min}$ serves as the tiebreaker. These yield a symbolic comparison key
\begin{equation}
\kappa(\hat{e}) = \bigl(|\mathcal{I}_{\not\sim}|,\; K,\; -H_{(1)},\; -H_{(2)},\; \dots,\; -H_{(n)}\bigr),
\label{eq:key}
\end{equation}
where $H_{(1)} \leq H_{(2)} \leq \cdots$ are the sorted specificity values of the non-overlapping goal states. Lexicographic minimisation of $\kappa$ faithfully preserves the ordering induced by~\eqref{eq:objective}, using only small computable integers and entirely bypassing the astronomically large intermediate quantities. Next, we present the experiment results.

\begin{figure}[t]
\centering
\includegraphics[width=0.9\columnwidth]{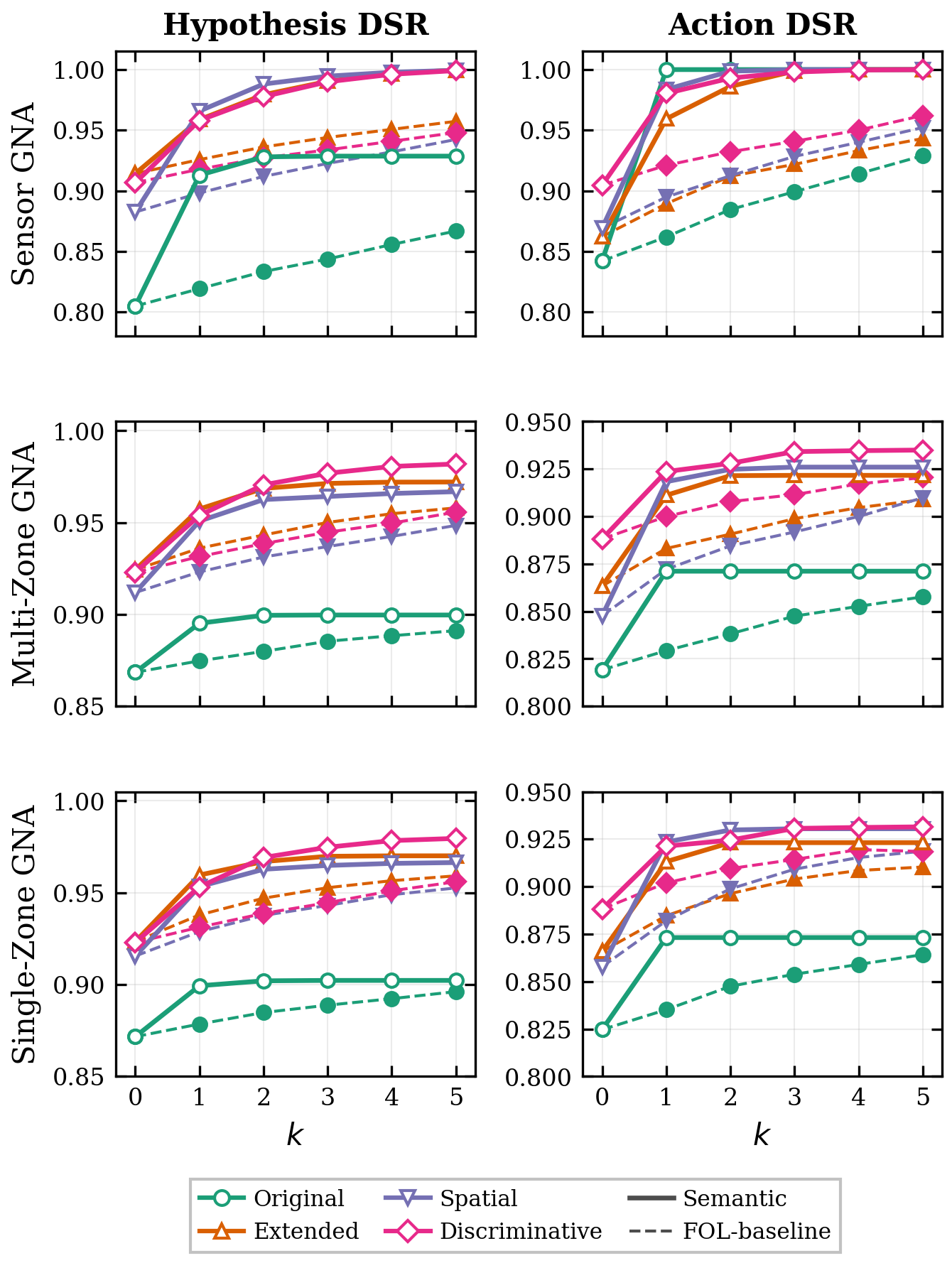}
\caption{Driving Simulator Experiment Results}
\label{fig:curves}
\vspace{-0.5cm}
\end{figure}

\section{Experimental Results}

We evaluate communication strategies in \textsc{LogiCity}~\cite{Li2024LogiCityAN}, a neuro-symbolic urban traffic simulator on a $241\times241$ grid, where cars and pedestrians navigate according to FOL traffic rules evaluated by a Z3 solver. The code for the experiments is open-sourced\footnote{https://github.com/ahmetfsaz/LogiCity}. Each agent makes observations within a field of view (FOV) of radius $r_{\text{fov}}$ and is affected by entities within a larger vicinity of radius $r_{\text{vic}}$. All agents always have access to observations in their FOV; and semantic communication provides additional observations from the surrounding vicinity through an edge server. Each observation is in the form of a single FOL expression. Agents evaluate several hypotheses based on their observations to choose from a set of actions $\{Stop, Slow, Fast, Normal\}$. There are multiple hypotheses that can result in the same action. We evaluate across four rule sets: the original LogiCity set (12 rules), an extended set with more hypotheses (48 rules), a spatial set emphasizing spatially-dependent hypotheses (30 rules), and a discriminative set designed to maximize entity diversity among hypotheses (30 rules).

Performance is measured against a full-information (FI) baseline in which the agent observes all FOV and vicinity entities. \emph{Hypothesis Decision Success Rate} (H-DSR) is the fraction of hypotheses evaluations that agree with the FI-baseline as a result of semantic communication. \emph{Action Decision Success Rate} (A-DSR) is the fraction of time steps for which the decided action (Stop, Slow, Fast, or Normal) from semantic communication matches the FI-baseline.

We compare the semantic selection strategy, minimizing~\eqref{eq:objective}, against a FOL-baseline that selects $k$ FOL expressions uniformly at random, under three architectures. In the \emph{Sensor Global Navigation Assistant (GNA)} setting, roadside sensors distributed across the grid provide direct access to all entities in each agent's vicinity, and the GNA selects $k$ expressions to transmit. In the \emph{Single-Zone GNA} setting, each vehicle uploads its FOV observations to a centralized assistant, which aggregates, filters for vicinity relevance, and returns $k$ FOL expressions to each agent. In the \emph{Multi-Zone Global Navigation Assistant (GNA)} setting, the grid is partitioned into non-overlapping zones, each served by a GNA that aggregates uplink transmissions from vehicles within its zone, filters for relevance, and transmits $k$ expressions to each agent within its zone.

Fig.~\ref{fig:curves} summarizes results for $r_{\text{fov}}{=}5$, $r_{\text{vic}}{=}25$, 20 cars, 8 pedestrians across all four rule sets. The most striking finding is that semantic selection dominates in the low-rate regime. Under sensor GNA, a single semantically chosen entity ($k{=}1$) recovers nearly all decision-relevant information: on the best configuration (spatial rules), A-DSR jumps from $0.869$ to $0.984$, reaching $1.0$ by $k{=}3$. The FOL-baseline, by contrast, improves only linearly---each additional random FOL expression has roughly equal chance of being useful---and reaches only $0.952$ at $k{=}5$. The intuition is clear: semantic objective identifies the FOL expression that resolves the most hypothesis uncertainty first, whereas the FOL-baseline must accumulate transmissions to stumble upon the critical observation.

Under the GNA architectures, semantic selection still outperforms FOL-baseline, but gains are more modest due to a structural bottleneck. The GNA pool contains only observations that are already observed by some vehicle, creating a Goldilocks problem: when FOV is too small (e.g., FOV${=}5$), the uplink GNA pool is too sparse for any strategy to differentiate; when too large (e.g., FOV${=}15$), agents already see most vicinity entities directly, leaving little for the GNA to communicate. A viable range exists around FOV${=}7$--$10$. Single-zone versus multi-zone organization makes almost no difference ($<0.3$\,pp), confirming that collective FOV coverage of all agents---not zone fragmentation---is the bottleneck, though multi-zone partitioning further starves local pools, which is why single-zone GNA performs slightly better. Nevertheless, semantic at $k_2{=}1$ matches random at $k_2{=}5$ in most experiments, preserving a $2.5$--$5\times$ bandwidth advantage.

Additional experiments with FOV${=}5$--$20$, cars${=}10$--$30$, pedestrians${=}4$--$12$ across all rule sets yield consistent results. Across all 17 configurations, semantic at $k{=}1$ matches or exceeds random at $k{=}5$---a $5\times$ bandwidth saving---though the advantage saturates around $k{=}3$. The variations point to a single conclusion: semantic selection helps most where the decision problem is hardest. Shrinking FOV widens the ring of unseen-but-relevant entities, growing the A-DSR advantage; increasing pedestrian density activates more entity-type-triggered hypotheses that FOL-baseline struggles to cover, widening the H-DSR gap; sparser environments make each entity more consequential, nearly tripling the action-level gap. This is captured by the correlation between baseline (no-communication) A-DSR and semantic advantage at $k{=}3$: $r{=}-0.974$---the benefit is almost perfectly predictable from how much the agent struggles without communication. Across every experiment, architecture, $k>0$, and metric, semantic selection never underperforms random, with the largest advantages at the tightest budgets.

\vspace{-0.15cm}

\bibliographystyle{unsrt} 
\bibliography{refs} 

@INPROCEEDINGS{swami,
  author={Bao, Jie and Basu, Prithwish and Dean, Mike and Partridge, Craig and Swami, Ananthram and Leland, Will and Hendler, James A.},
  booktitle={2011 IEEE Network Science Workshop}, 
  title={Towards a theory of semantic communication}, 
  year={2011},
  volume={},
  number={},
  pages={110-117},
  doi={10.1109/NSW.2011.6004632}}

@article{c1,
	author = {Rudolf Carnap},
	journal = {Philosophy},
	number = {106},
	pages = {272--273},
	publisher = {Cambridge University Press},
	title = {The Continuum of Inductive Methods},
	volume = {28},
	year = {1953}
}

@inproceedings{c2,
  title={An outline of a theory of semantic information},
  author={Rudolf Carnap and Yehoshua Bar-Hillel},
  year={1952},
  url={https://api.semanticscholar.org/CorpusID:11969100}
}

@article{c4,
  title={Relational hypotheses and inductive inference},
  author={Risto Hilpinen},
  journal={Synthese},
  year={1971},
  volume={23},
  pages={266-286},
  url={https://api.semanticscholar.org/CorpusID:46963142}
}

@article{c9,
  title={Distributive Normal Forms in First-Order Logic},
  author={Jaakko Hintikka},
  journal={Studies in logic and the foundations of mathematics},
  year={1965},
  volume={40},
  pages={48-91},
  url={https://api.semanticscholar.org/CorpusID:118372068}
}

@article{c10,
  title={A Two-Dimensional Continuum of Inductive Methods*},
  author={Jaakko Hintikka},
  journal={Studies in logic and the foundations of mathematics},
  year={1966},
  volume={43},
  pages={113-132},
  url={https://api.semanticscholar.org/CorpusID:118253044}
}

@article{Xie2020DeepLE,
  title={Deep Learning Enabled Semantic Communication Systems},
  author={Huiqiang Xie and Zhijin Qin and Geoffrey Y. Li and Biing-Hwang Juang},
  journal={IEEE Transactions on Signal Processing},
  year={2020},
  volume={69},
  pages={2663-2675},
  url={https://api.semanticscholar.org/CorpusID:219792180}
}

@mastersthesis{nelte,
    author = {Nelte, Karen},
    title = {Formulas of First-Order Logic in Distributive Normal Form},
    school = {University of Cape Town},
    year = {1997}
}

@book{carnap_logical_1962,
	address = {Chicago},
	edition = {2nd},
	title = {The logical foundation of probability},
	publisher = {The University of Chicago Press},
	author = {Carnap, Rudolf},
	year = {1962},
}

@article{Yang2023SwinJSCCTS,
  title={SwinJSCC: Taming Swin Transformer for Deep Joint Source-Channel Coding},
  author={Ke Yang and Sixian Wang and Jincheng Dai and Xiaoqi Qin and Kai Niu and Ping Zhang},
  journal={IEEE Transactions on Cognitive Communications and Networking},
  year={2023},
  volume={11},
  pages={90-104},
  url={https://api.semanticscholar.org/CorpusID:261031211}
}

@article{Niu2024AMT,
  title={A Mathematical Theory of Semantic Communication},
  author={Kai Niu and Ping Zhang},
  journal={ArXiv},
  year={2024},
  volume={abs/2401.13387},
  url={https://api.semanticscholar.org/CorpusID:267199848}
}

@article{Shao2022ATO,
  title={A Theory of Semantic Communication},
  author={Yulin Shao and Qingqing Cao and Deniz G{\"u}nd{\"u}z},
  journal={IEEE Transactions on Mobile Computing},
  year={2022},
  volume={23},
  pages={12211-12228},
  url={https://api.semanticscholar.org/CorpusID:254247048}
}

@article{Niu2025RateDistortionPerceptionTI,
  title={Rate–Distortion–Perception Trade-Off in Information Theory, Generative Models, and Intelligent Communications},
  author={Xueyan Niu and Bo Bai and Nian Guo and Weixi Zhang and Wei Han},
  journal={Entropy},
  year={2025},
  volume={27},
  url={https://api.semanticscholar.org/CorpusID:277527222}
}

@article{Saz2025AnalysisOS,
  title={Analysis of Semantic Communication for Logic-based Hypothesis Deduction},
  author={Ahmet Faruk Saz and Siheng Xiong and Faramarz Fekri},
  journal={GLOBECOM 2025 - 2025 IEEE Global Communications Conference},
  year={2025},
  pages={5701-5707},
  url={https://api.semanticscholar.org/CorpusID:280985028}
}

@article{Saz2024LossySC,
  title={Lossy Semantic Communication for the Logical Deduction of the State of the World},
  author={Ahmet Faruk Saz and Siheng Xiong and Faramarz Fekri},
  journal={2025 IEEE Wireless Communications and Networking Conference (WCNC)},
  year={2024},
  pages={1-6},
  url={https://api.semanticscholar.org/CorpusID:273025733}
}

@INPROCEEDINGS{Saz2024OnTT,
  author={Saz, Ahmet Faruk and Xiong, Siheng and Fekri, Faramarz},
  booktitle={2026 IEEE Wireless Communications and Networking Conference (WCNC)}, 
  title={On the Theory of Semantic Information and Communication for Logical Inference}, 
  year={2026},
  volume={},
  number={},
  pages={1-6},
  doi={10.1109/WCNC65185.2026.11555338}}

@inproceedings{Krygin2024PerformingFQ,
  title={Performing First-Order-Logic Queries Over RDF Data: Interpreter Versus Compiler to Apache Jena Rules},
  author={Andrey Krygin and Pavel Karpenko and Oleg Sychev},
  booktitle={International Conference on Novelties in Intelligent Digital Systems},
  year={2024},
  url={https://api.semanticscholar.org/CorpusID:274233387}
}

@article{Li2024LogiCityAN,
  title={LogiCity: Advancing Neuro-Symbolic AI with Abstract Urban Simulation},
  author={Bowen Li and Zhaoyu Li and Qiwei Du and Jinqi Luo and Wenshan Wang and Yaqi Xie and Simon Stepputtis and Chen Wang and Katia P. Sycara and Pradeep Kumar Ravikumar and Alexander Gray and Xujie Si and Sebastian A. Scherer},
  journal={ArXiv},
  year={2024},
  volume={abs/2411.00773},
  url={https://api.semanticscholar.org/CorpusID:273798388}
}

@inproceedings{Niiniluoto2011TheDO,
  title={The Development of the Hintikka Program},
  author={Ilkka Niiniluoto},
  booktitle={Inductive Logic},
  year={2011},
  url={https://api.semanticscholar.org/CorpusID:8156343}
}

\newpage

\appendices

\section{Inductive Logical Probability as a Foundation for Semantic Information}
\label{sec:logical_probability}

The communication framework developed in this work rests on an inductive logical probability measure assigned to the semantic structures of a first-order language $\mathcal{L}$. This measure satisfies Kolmogorov's axioms and, through the rules of logic, determines the probabilities of all sentences expressible in $\mathcal{L}$. In this sense, probability is not introduced as a frequency over directly observed events, but as a principled measure over logically described possibilities. The empirical component of the framework enters through the choice of likelihoods and the use of Bayesian updating, so that the resulting probability assignments are both logically organized and empirically informed. In this section, we situate our approach within the tradition of logical probability initiated by Carnap~\cite{carnap_logical_1962} and extended by Hintikka~\cite{Niiniluoto2011TheDO}, clarify the specific modelling choices we adopt, and identify the points at which our formulation departs from the classical programme.

Logical probability, as conceived by Carnap and Keynes, defines probability as the degree of support or partial entailment that a body of evidence $e$ provides for a hypothesis $h$, denoted $c(h, e)$. This quantity expresses an epistemic, analytic relationship between propositions rather than a physical frequency in the world. A foundational consequence of this definition is that all probabilities are inherently conditional: a probability statement is meaningful only relative to some body of evidence or background assumptions. This feature attracted sustained criticism from both the frequentist and Bayesian traditions, and alternative programmes came to dominate the discourse on the foundations of probability. Nevertheless, for the purpose of quantifying semantic information, the logical interpretation proves especially fruitful, a view supported by the classical semantic information theory of Carnap and Bar-Hillel~\cite{c2} and subsequent work on semantic communication~\cite{swami}.

Our use of logical probability is closely aligned with an objective Bayesian interpretation. The uncertainty of interest is epistemic: it arises from an agent's incomplete knowledge of the true state of the world. In the construction adopted here, world states are modelled through \emph{constituents}---the maximally specific and complete descriptions of the universe expressible in~$\mathcal{L}$, corresponding to Hintikka's distributive normal forms---and a probability measure is ascribed to their set. The core intuition is analogous to reasoning about a well-shuffled deck of 52 cards: the probability that the top card is black is $26/52 = 0.5$, not because of any observed frequency of draws, but because that is the proportion of the state space satisfying the hypothesis. In our framework, the role of the deck is played by the set of constituents, and hypotheses are evaluated by the proportion of constituents in which they hold. More generally, due to the hierarchical structure of the logical language, once the probabilities of constituents are fixed, the probability of any sentence follows from the set of constituents it satisfies; logical probability thus reflects an agent's epistemic uncertainty over world states, which in turn determines the probability of every sentence in the language. Evidence updates the prior over constituents through Bayes' rule, producing posterior probabilities that alter the degree of support each world state receives. Since any sentence can be expressed as a disjunction of constituents, this update propagates systematically to every sentence of interest. Furthermore, logical probability is objective in character: the relevant probabilities are determined by the logical structure of the language together with the adopted inductive method, and any rational agent aware of both can \emph{objectively} reproduce the same assignments. Given the same language framework and inductive method, every rational agent therefore arrives at the same degree of confirmation for any evidence--hypothesis pair.

Our formulation adopts this Carnapian objective Bayesian perspective but departs from the classical programme on logical probability in two important respects. The first concerns the treatment of evidence. Carnap drew a sharp distinction between the \emph{logical problem of confirmation}---determining the degree of support that the logical structure of two sentences yields---and the \emph{methodological problems of confirmation}, which concern practical matters such as how to construct experiments, carry out observations, and obtain evidential results~\cite{carnap_logical_1962}. His confirmation function $c(h, e)$ was designed to address only the former: to determine the degree to which $e$ confirms $h$, one need not know whether $e$ is true or false, or on what physical basis it was obtained; all that is required is a logical analysis of the meanings of the two sentences. Accordingly, in both Carnap's original framework and Hintikka's later work, evidence is treated as a fixed description of observed facts, and the physical process by which it arises is set aside as a methodological concern outside the scope of the logical programme.

We do not adopt this separation in full. Instead, we bring the origin of evidence into the formalism by treating it as a random quantity whose marginal probability is obtained by marginalising over the set of world states:
\begin{equation}
p(e) = \sum_{C \in \mathcal{C}} p(e \mid C)\,\pi(C),
\end{equation}
where $\pi(C)$ is the prior over constituents and $p(e \mid C)$ is the likelihood of evidence under constituent $C$. The epistemic uncertainty over world states thus carries over to the marginal probability of evidence. Importantly, we do not introduce a separate stochastic model for evidence generation or separately model the inherent uncertainty of the physical world; rather, all uncertainty is attributed to uncertainty regarding world states, and the marginal probability of evidence is obtained by averaging likelihoods with respect to the prior over those states. Once evidence is observed and the corresponding sentence is formulated, the degree of confirmation $c(h, e)$ remains a purely analytic, framework-relative relationship between evidence and hypothesis---precisely as in Carnap's logical programme. The departure lies in what happens \emph{before} observation: by assigning evidence a well-defined marginal probability within the same probabilistic framework, we account for the uncertainty of evidence generation without relegating it to a separate methodological domain. Consequently, the marginal probability $p(e)$ appears explicitly in our formulae for semantic content, entropy, and mutual information---a feature absent from the classical formulations of Carnap and Hintikka, where evidence enters only through conditioning.

The second departure concerns the determination of the inductive method itself. Carnap initially insisted that inductive confirmation should be determined independently of empirical and methodological concerns---that is, independently of extra-logical factors. His use of symmetry and exchangeability assumptions, his appeal to De Finetti's representation theorem, and his final "Basic System" grounded in geometrical considerations over quality spaces all exemplify this ambition: to arrive at an objective, rational, and purely analytic method of assigning confirmation without recourse to empirical input. Hintikka's subsequent development replaced these purely analytic principles with more liberal formulations that explicitly admit extra-logical factors, as documented by Niiniluoto~\cite{Niiniluoto2011TheDO}. Our framework is compatible with these later developments: while symmetry principles over the language may still determine priors over world states, likelihoods may legitimately incorporate empirical considerations, and, depending on the application, even the priors themselves may be informed as such. We do not require the inductive function, the prior, or the likelihood model to be fixed solely by logic. Instead, these components may be shaped by empirical structure, domain knowledge, and modelling assumptions. The role of the logical language in our framework is therefore not to eliminate empirical input, but to impose a coherent semantic organisation on probabilistic reasoning: once probabilities are assigned at the level of world states and likelihoods are specified, the logical structure of the language systematically propagates these assignments to all sentences of interest. We adopt Carnap's symmetry assumptions---specifically, the uniform prior over constituents---as a default in the absence of domain-specific reasons to deviate, but regard this as a modelling choice rather than a logical necessity.

Finally, this interpretation clarifies the relation between epistemic and aleatoric uncertainty in our setting. Aleatoric uncertainty concerns irreducible randomness in outcomes, whereas epistemic uncertainty arises from incomplete knowledge of the relevant system, mechanism, or probability law. Our framework is centred on epistemic uncertainty: the agent does not know which world state is actual, and this ignorance induces uncertainty about hypotheses and evidence alike. By grounding the probability assignments in logical descriptions of world states and by adopting symmetry principles unless domain-specific reasons suggest otherwise, the resulting framework remains closer to objective Bayesianism. The present approach may thus be understood as a logically structured, empirically informed, and epistemically grounded probabilistic foundation for semantic communication for logical reasoning.

\section{Goal-Oriented Semantic Communication Algorithm}
 
The entity selection objective is
\begin{equation}
    \mathcal{F}(\hat{E}) = \sum_{i=1}^{M} p(\hat{\mathbf{e}}) \cdot p(\Gamma_i \mid \hat{\mathbf{e}}) \cdot \bigl(1 - p(\Gamma_i \mid \hat{\mathbf{e}})\bigr),
\end{equation}
where each summand is the evidence-weighted Bernoulli variance of hypothesis~$\Gamma_i$, maximised at $p=0.5$ (full uncertainty) and zero at $p\in\{0,1\}$ (full certainty).
 
\subsection*{Closed-Form Substitution}
 
From the constituent-based probability model (Theorems~1--2), the evidence probability and the conditional hypothesis probability for a non-overlapping hypothesis are
\begin{equation}
    p(\hat{\mathbf{e}}) = 1 - v, \qquad p(\Gamma_i \mid \hat{\mathbf{e}}) = \frac{1 - u_i}{1 - v},
    \label{eq:closed_forms}
\end{equation}
where $v \triangleq 2^{-\alpha}$, $u_i \triangleq 2^{-\gamma_i}$, $\alpha \triangleq 2^{Q-K}$, $\gamma_i \triangleq 2^{Q-K} - 2^{Q-K-H_i}$, $Q = 2^T$ is the total number of Q-sentences, $K$ is the number of distinct Q-sentences in the evidence, and $H_i = 2^{T-Z_i}$ is the number of Q-sentences compatible with hypothesis~$\Gamma_i$ (which fixes $Z_i$ predicate slots). If any evidence Q-sentence already satisfies~$\Gamma_i$ (the overlap case), then $p(\Gamma_i \mid \hat{\mathbf{e}}) = 1$ and the corresponding term vanishes. For every non-overlapping hypothesis, the complement is
\begin{equation}
    1 - p(\Gamma_i \mid \hat{\mathbf{e}}) = 1 - \frac{1 - u_i}{1 - v} = \frac{(1-v) - (1-u_i)}{1-v} = \frac{u_i - v}{1 - v}.
    \label{eq:complement}
\end{equation}
Substituting~\eqref{eq:closed_forms} and~\eqref{eq:complement} into the per-hypothesis term gives
\begin{align}
    \mathcal{F}_i &= (1 - v) \cdot \frac{1 - u_i}{1 - v} \cdot \frac{u_i - v}{1 - v} \notag \\
    &= \frac{(1 - u_i)(u_i - v)}{1 - v}.
    \label{eq:simplified_term}
\end{align}
Note that $0 < v < u_i < 1$ (since $\gamma_i < \alpha$), so every factor is strictly positive.
 
\subsection*{Numerical Intractability}
 
For our language parameters ($T = 34$, hence $Q = 2^{34} \approx 1.7\times10^{10}$), the exponents $\alpha = 2^{Q-K}$ and $\gamma_i = 2^{Q-K} - 2^{Q-K-H_i}$ are integers with billions of digits. Consequently, $v = 2^{-\alpha}$ and $u_i = 2^{-\gamma_i}$ underflow to zero in IEEE~754 arithmetic, and every candidate subset evaluates to $\mathcal{F} = 0$, making direct computation infeasible.
 
\subsection*{Asymptotic Reduction}
 
Since $\alpha, \gamma_i \gg 1$, both $u_i \to 0^+$ and $v \to 0^+$. In this regime, \eqref{eq:simplified_term} simplifies as
\begin{equation}
    \mathcal{F}_i = \frac{(1 - u_i)(u_i - v)}{1 - v} \;\approx\; \frac{u_i - v}{1} \;\approx\; u_i = 2^{-\gamma_i},
    \label{eq:asymptotic}
\end{equation}
where the second step uses $v \ll u_i$ (since $\alpha \gg \gamma_i$). The total objective becomes
\begin{equation}
    \mathcal{F}(\hat{E}) = \sum_{i \in \mathcal{I}_{\not\sim}} 2^{-\gamma_i} \;\approx\; 2^{-\gamma_{\min}}, \qquad \gamma_{\min} \triangleq \min_{i \in \mathcal{I}_{\not\sim}} \gamma_i,
    \label{eq:dominated}
\end{equation}
since all terms are positive and the term with the smallest exponent dominates the sum exponentially. Minimising $\mathcal{F}$ therefore reduces to maximising~$\gamma_{\min}$.
 
\subsection*{Decomposition into Independent Effects}
 
Factoring $\gamma_i$ reveals
\begin{equation}
    \gamma_i = 2^{Q-K} - 2^{Q-K-H_i} = 2^{Q-K-H_i}\bigl(2^{H_i} - 1\bigr).
    \label{eq:gamma_factor}
\end{equation}
Comparing $\gamma_{\min}$ across two candidate subsets $\hat{E}_a$ (with $K_a$ distinct Q-sentences) and $\hat{E}_b$ (with $K_b$) yields two hierarchically ordered effects:
 
\emph{1) Evidence diversity ($K$):} If $K_a < K_b$, the leading term $2^{Q-K_a}$ exceeds $2^{Q-K_b}$ by a factor of $2^{K_b - K_a}$, which propagates \emph{doubly exponentially} through~\eqref{eq:gamma_factor} to increase every $\gamma_i$ simultaneously. This effect dominates all others.
 
\emph{2) Hypothesis specificity ($H_{\min}$):} When $K_a = K_b$, the bottleneck hypothesis (i.e., the one with the smallest $H_i$) determines $\gamma_{\min}$, since $2^{Q-K-H_i}$ is largest---and hence $\gamma_i$ smallest---for the smallest~$H_i$. Maximising $H_{\min} = \min_{i \in \mathcal{I}_{\not\sim}} H_i$ then maximises~$\gamma_{\min}$.
 
\subsection*{Symbolic Comparison Key}
 
These observations yield the lexicographic key
\begin{equation}
    \kappa(\hat{E}) = \bigl(|\mathcal{I}_{\not\sim}|,\; K,\; -H_{(1)},\; -H_{(2)},\; \dots,\; -H_{(n)}\bigr),
    \label{eq:key}
\end{equation}
where $H_{(1)} \leq H_{(2)} \leq \cdots$ are the sorted specificity values of the non-overlapping hypotheses. Lexicographic minimisation of $\kappa$ preserves the ordering induced by~\eqref{eq:objective}: the first component counts positive terms in the sum, the second captures the doubly-exponential effect of $K$, and the remaining components resolve ties via $H_{\min}$ and beyond. Crucially, $\kappa$ involves only small integers ($|\mathcal{I}_{\not\sim}|$, $K$, and the $H_i$ values), entirely bypassing the astronomically large intermediate quantities and enabling exact, tractable optimisation.

Algorithm~\ref{alg:semantic_selection} operationalises the complete procedure.

\begin{algorithm}
\caption{Goal-Oriented Semantic Entity Selection}
\label{alg:semantic_selection}
\begin{algorithmic}[1]
\REQUIRE Vicinity entities $E = \{e_1, \dots, e_n\}$ with grounded predicates for each pair $(x_0, e_j)$; slot map $\sigma$; hypotheses $\{\Gamma_1, \dots, \Gamma_M\}$ with fixed-slot sets $\{(s,v)\}$; predicate slot count $T$; budget $k$
\ENSURE Optimal subset $\hat{E}^*$
\medskip
\STATE \textbf{// Phase 1: Compute Q-sentences for all entities}
\FOR{$j = 1$ \TO $n$}
    \STATE $q_j \leftarrow \mathbf{0} \in \{0,1\}^T$
    \FOR{each grounded predicate $(c, P, val)$ of pair $(x_0, e_j)$}
        \IF{$(c, P) \in \operatorname{dom}(\sigma)$}
            \STATE $q_j[\sigma(c, P)] \leftarrow val$
        \ENDIF
    \ENDFOR
\ENDFOR
\medskip
\STATE \textbf{// Phase 2: Exhaustive subset evaluation}
\IF{$n \leq k$}
    \RETURN $E$ \hfill $\triangleright$ Pool does not exceed budget
\ENDIF
\STATE $\kappa^* \leftarrow (+\infty, \dots)$; \quad $\hat{E}^* \leftarrow \emptyset$
\FOR{each $\hat{E} \subseteq E$ with $|\hat{E}| = k$}
    \STATE $\mathcal{S} \leftarrow \{q_j : e_j \in \hat{E}\}$ \hfill $\triangleright$ Distinct Q-sentences
    \STATE $K \leftarrow |\mathcal{S}|$
    \medskip
    \STATE \textbf{// Phase 3: Symbolic comparison key}
    \STATE $\mathcal{H} \leftarrow \emptyset$ \hfill $\triangleright$ $H_i$ values for non-overlapping hypotheses
    \FOR{$i = 1$ \TO $M$}
        \STATE $\textit{overlap} \leftarrow \FALSE$
        \FOR{each $q \in \mathcal{S}$}
            \IF{$\forall\, (s, v) \in \Gamma_i : q[s] = v$}
                \STATE $\textit{overlap} \leftarrow \TRUE$; \textbf{break}
                \hfill $\triangleright$ $p(\Gamma_i \mid \hat{\mathbf{e}}) = 1 \Rightarrow \mathcal{F}_i = 0$
            \ENDIF
        \ENDFOR
        \IF{$\neg\,\textit{overlap}$}
            \STATE $Z_i \leftarrow |\Gamma_i|$ \hfill $\triangleright$ Number of fixed predicate slots
            \STATE $H_i \leftarrow 2^{T - Z_i}$ \hfill $\triangleright$ Hypothesis-compatible Q-sentences
            \STATE $\mathcal{H} \leftarrow \mathcal{H} \cup \{H_i\}$
        \ENDIF
    \ENDFOR
    \STATE Sort $\mathcal{H}$ ascending: $H_{(1)} \leq H_{(2)} \leq \cdots \leq H_{(|\mathcal{H}|)}$
    \STATE $\kappa \leftarrow \bigl(|\mathcal{H}|,\; K,\; {-H_{(1)}},\; {-H_{(2)}},\; \dots,\; {-H_{(|\mathcal{H}|)}}\bigr)$
    \medskip
    \STATE \textbf{// Phase 4: Lexicographic comparison}
    \IF{$\kappa <_{\text{lex}} \kappa^*$}
        \STATE $\kappa^* \leftarrow \kappa$; \quad $\hat{E}^* \leftarrow \hat{E}$
    \ENDIF
\ENDFOR
\RETURN $\hat{E}^*$
\end{algorithmic}
\end{algorithm}

\section{Models for Representation of Structure and Probabilities of the Logical Language $\Language$}

\subsection{Probabilistic Models over the Language Structure}

The layered structure of the language $\Language$ presented in this manuscript, $\mathcal{P} \to \mathcal{Q} \to C_j \to C^w$, is a logical construction: it defines the semantic building blocks. These blocks do not presuppose any particular probabilistic model. Unlike many natural random processes that possess an intrinsic generative hierarchy—most observational data arise from hidden generative processes—the logical language does not come equipped with a canonical data-generating mechanism. In other words, neither the structure nor any assignment of probabilities arises naturally. Rather, interpreting the language through a particular structural model and probability assignment is an artificial—but highly productive—imposition, and more than one such interpretation is possible.

The hierarchical Bayesian reading (constituents generate attributive constituents, which generate Q-relationships, which generate observed evidence) is one of several admissible probabilistic characterizations. Alternatives include: treating the evidence as exact logical constraints with an indicator likelihood; computing likelihoods via matrix eigenvalues over transition graphs; or defining likelihoods through variational bounds without committing to a specific generative story. Each characterization imposes its own statistical model on the same logical scaffolding, and the choice among them is a modelling decision driven by the application, the available evidence structure, and the desired tradeoff between expressiveness and tractability.

All characterization share the fundamental principle that constituent probabilities determine all other sentence probabilities:
\begin{align}
p(C^w \wedge e) &= p(C^w)\,p(e \mid C^w), \\
p(e) &= \textstyle\sum_i p(C^{(i)})\,p(e \mid C^{(i)}), \\
p(h \mid e) &= \frac{\sum_{i \in I_h} p(C^{(i)})\,p(e \mid C^{(i)})}{p(e)},
\end{align}
where $h$ is any sentence and $I_h$ indexes the constituents satisfying $h$. They differ in how $p(e \mid C^w)$ is defined and computed.

\subsection{Interpretation as a Hierarchical Bayesian Model}

As stated, one probabilistic reading of the language structure interprets it as a hierarchical Bayesian model with a two-stage generative process: (1)~the true constituent determines which kinds of individuals (attributive constituents) are permitted; (2)~individuals are sampled from the permitted kinds, each exhibiting Q-relationships determined by their attributive constituent. Evidence consists of observed Q-relationships, which are typically partial observations of the full attributive constituent of each individual.

\subsubsection{Connection to Finite Mixture Models}

In this interpretation, the constituent $C^w$ functions as a latent model index, $p(C^w)$ is a mixing weight, and $p(e \mid C^w)$ is a component-specific likelihood. The marginal likelihood of the evidence takes the standard finite-mixture form:
\[
p(e) = \sum_{w} p(C^w)\,p(e \mid C^w) = \sum_{w} \pi_w\,f_w(e),
\]
where $\pi_w = p(C^w)$ and $f_w(e) = p(e \mid C^w)$. Different constituents induce different predictive laws, and the final predictive probability is obtained by Bayesian model averaging:
\[
p(X_{n+1} = j \mid e) = \sum_w p(C^w \mid e)\,p(X_{n+1} = j \mid e, C^w).
\]
Hintikka explicitly frames this via Good's distinction: a \emph{Type~I distribution} $p(\cdot \mid C^w)$ is the predictive law once the constituent is fixed; a \emph{Type~II distribution} $p(C^w)$ governs uncertainty over which predictive law applies. Thus, the framework assigns probabilities not merely to outcomes, but to \emph{ways of predicting outcomes}.

Hilpinen's extension makes the hierarchy even more explicit by introducing a second latent layer---the E-distribution $D_\nu$---giving:
\[
p(e) = \sum_{w,\nu} p(C^w)\,p(D_\nu \mid C^w)\,p(e \mid C^w, D_\nu).
\]
This is a two-level latent-variable model: first a global structure is drawn ($C^w$), then a local assignment of individuals to types ($D_\nu$), then the relational evidence is generated conditionally. In modern terminology, this is a hierarchical finite mixture with latent class assignments at the individual level. To elaborate, Hilpinen's dyadic model is a hierarchical finite mixture with three layers of structure:

\begin{enumerate}
\item \textbf{Global layer:} A constituent $C^w$ is drawn from a prior $p(C^w)$, determining which attributive constituent types are permitted and their mixing weights $\pi_j^{(w)} = p(T_i = j \mid C^w)$.
\item \textbf{Local layer:} Each observed individual $a_i$ is independently assigned a latent (attributive constituent) type $T_i \in \{1, \dots, w\}$ drawn from $\pi^{(w)}$. The collection $(T_1, \dots, T_n)$ constitutes the E-distribution $D_\nu$.
\item \textbf{Data layer:} Relational evidence between pairs $(a_h, a_k)$ is generated conditionally on their types $(T_h, T_k)$, restricted to the $q_{T_h T_k}$ logically admissible Q-sentences.
\end{enumerate}

The full marginal likelihood is:
\[
p(e) = \sum_{w} p(C^w) \sum_{T_1, \dots, T_n} \prod_{i=1}^{n} p(T_i \mid C^w) \; p(e \mid T_1, \dots, T_n, C^w).
\]

The model is \emph{hierarchical} because the local mixing weights $\pi_j^{(w)}$ depend on the global latent variable $C^w$ (unlike a flat mixture where they are fixed parameters), and \emph{finite} because both layers involve discrete latent variables with finitely many values ($2^K$ constituents, $w^n$ possible type assignments). Logic enters through structural zeros: Q-sentences incompatible with a given type pair receive probability exactly zero. The closest modern analogues are stochastic block models (attributive types as communities, Q-sentences as edges), latent Dirichlet allocation (the world as a document, types as topics, Q-observations as words), and latent class models with uncertain class structure.

\subsubsection{Markov Chain and Conditional Independence Structure}
\label{sec:markov}

The generative hierarchy induces a natural Markov chain:
\[
C^w \;\longrightarrow\; D_\nu \;\longrightarrow\; e.
\]
Under this chain, evidence and constituent are conditionally independent given the E-distribution: $C^w \perp e \mid D_\nu$, so $p(e \mid C^w, D_\nu) = p(e \mid D_\nu)$ whenever the E-distribution fully specifies the local structure. A richer chain arises when attributive constituents and Q-sentences are separated as distinct latent layers:
\[
C^w \;\longrightarrow\; \mathrm{Ct}_j \;\longrightarrow\; Q_i \;\longrightarrow\; e.
\]
Each link carries a conditional independence property: once $\mathrm{Ct}_j$ is known, the constituent $C^w$ provides no additional information about $Q_i$ or $e$. Formally, $p(C^w \mid \mathrm{Ct}_j, Q_i, e) = p(C^w \mid \mathrm{Ct}_j)$. These conditional independence relations are what make it possible to decompose the posterior via the law of total probability:
\begin{align}
\nonumber p(C^w \mid e) &= \sum_j p(\mathrm{Ct}_j \mid e)\,p(C^w \mid \mathrm{Ct}_j) \\
\nonumber  &= \sum_{i,j} p(Q_i \mid e)\,p(\mathrm{Ct}_j \mid Q_i)\,p(C^w \mid \mathrm{Ct}_j).
\end{align}
These simplifications are valid only under the assumed Markov structure. Using an externally fixed $p(Q_i \mid e)$---e.g., from equal splitting---is a serviceable approximation but does not coincide with the fully coherent Bayesian posterior under the generative chain.

\begin{remark}
Specifying $p(Q_i \mid \mathrm{Ct}_j)$ requires a sampling semantics (e.g., sampling a random pair, or an ego individual and then a partner). An attributive constituent $\mathrm{Ct}_j(x)$ specifies which Q-sentences are existentially realised for an individual, but not a probability distribution over them. Without an explicit sampling story, the transition $\mathrm{Ct}_j \to Q_i$ is logically meaningful but probabilistically incomplete.
\end{remark}

\subsection{Prior Over Constituents}

\subsubsection{The Hintikka--Hilpinen Prior}

Constituents with the same width $w$ are taken to be equiprobable. The prior for a particular constituent $C_w^{(i)}$ of width $w$ is:
\[
p(C_w^{(i)}) = \frac{p(\mathcal{C}^w)}{\binom{K}{w}}, \qquad p(\mathcal{C}^w) = \frac{\binom{K}{w}\,\pi(\alpha,\, w\lambda/K)}{\sum_{i=0}^{K} \binom{K}{i}\,\pi(\alpha,\, i\lambda/K)},
\]
where $\pi(\alpha, x) := \Gamma(\alpha + x)/\Gamma(\alpha)$ and $\alpha, \lambda$ are parameters of the inductive system, where $\alpha$ determines how fast inductive generalizations learn from evidence and $\lambda$ determines how fast singular inductive inferences learn from evidence. These parameters balance the impact of prior probability assignment with learning from evidence for different kinds of inductive inference.

The rationale derives from sequential prediction. Given a fixed constituent $C^w$, the probability that $\alpha$ successive individuals are all compatible with $C^w$ is a product of Dirichlet--multinomial predictive factors:
\[
\prod_{t=0}^{\alpha-1} \frac{t + w\lambda/K}{t + \lambda}.
\]
This is structurally identical to a P\'{o}lya urn: observed counts reinforce categories, and the prior pseudo-count $\lambda/K$ per category plays the role of initial balls. The predictive rule $p(\text{next is type } j \mid \text{past}) = (n_j + \lambda/K)/(n + \lambda)$ is a strict generalisation of Laplace's rule of succession and satisfies Johnson's count-sufficiency principle (dependence only on counts, not ordering).

\subsubsection{Direct Uniformity ($\alpha = 0$)}

Setting $\alpha = 0$ yields $\pi(0, x) = 1$ for all $x > 0$, so all $2^K$ constituents are equiprobable:
\[
p(C_w^{(i)}) = \frac{1}{2^K} \quad \text{for every constituent}, \qquad p(\mathcal{C}^w) = \frac{\binom{K}{w}}{2^K}.
\]

\subsection{Likelihood Models}

\subsubsection{Model~1: Monadic Likelihood (Hintikka)}

Applicable when the language contains only monadic predicates. Given evidence counts $n_1, \dots, n_c$ ($\sum n_j = n$), the likelihood under constituent $C^w$ is:
\[
p(e \mid C^w) = \frac{\prod_{j=1}^{c} \pi(n_j,\, \lambda/w)}{\pi(n,\, \lambda)}, \qquad \pi(n_j, \lambda/w) = \frac{\Gamma(n_j + \lambda/w)}{\Gamma(\lambda/w)}.
\]
This is a Dirichlet--multinomial characteristic function. Under the symmetric Dirichlet prior with $\alpha_j = \lambda/K$, the posterior predictive probability for the next observation is $(n_j + \lambda/K)/(n + \lambda)$---the same smoothed-count formula used in Bayesian categorical prediction.

\textbf{Special case} ($\alpha = 0$, $\lambda = w$):
\[
p(e \mid C^w) = \frac{(w-1)!}{(n+w-1)!} \prod_{j=1}^{c} (n_j)!,
\]
so the posterior depends on evidence only through $c$ (number of distinct Q-types observed) and the weight distribution $(n_j)$.

\subsubsection{Model~2: Dyadic Likelihood (Hilpinen)}

For dyadic languages, the likelihood decomposes into two stages via an E-distribution $D_\nu$, which assigns each observed individual to an attributive constituent class:
\[
p(C^w, D_\nu \mid e) = \frac{p(C^w)\,p(D_\nu \mid C^w)\,p(e \mid D_\nu, C^w)}{p(e)}.
\]

\textbf{First stage} --- attributive constituent assignment:
\[
p(D_\nu \mid C^w) = \frac{\Gamma(\lambda)}{\Gamma(n + \lambda)} \prod_{j=1}^{\nu} \frac{\Gamma(n_j + \lambda/w)}{\Gamma(\lambda/w)}.
\]
Any $D_\nu$ incompatible with $C^w$ receives zero probability. This term favours \emph{Ct-simple} hypotheses (fewer kinds of individuals).

\textbf{Second stage} --- relational likelihood. For each pair of individuals $(a_h, a_k)$ assigned to classes $\mathrm{Ct}_h$ and $\mathrm{Ct}_k$, the number of permissible Q-sentences between them is $q_{hk} = |A_{hk}|$:
\[
p(e \mid D_\nu, C^w) = \prod_{1 \leq h \leq k \leq \nu} \prod_{s=1}^{q_{hk}} \frac{(m_{hk}^{(s)})!\,(q_{hk} - 1)!}{(N_{hk} + q_{hk} - 1)!}.
\]
This term favours \emph{Q-simple} hypotheses (fewer admissible relational possibilities, i.e., smaller $q_{hk}$). The posterior $p(C^w \wedge D_\nu \mid e)$ thus balances two competing simplicity pressures: reducing the number of kinds of individuals (Ct-simplicity) and reducing the number of possible relations between them (Q-simplicity).

\begin{remark}[Constrained multinomials and structural zeros]
The quantities $q_{hk}$ function as local support sizes. Q-relations that are logically impossible under a given $(C^w, D_\nu)$ are structural zeros---assigned probability exactly zero by the model, not merely unobserved. This makes Hilpinen's framework a constrained multinomial model where logic determines which categories are allowed and statistics determines how probable the allowed categories are.
\end{remark}

\subsubsection{Model~3: Tuomela's Matrix Method}
\label{sec:tuomela}

Tuomela extends Hintikka's system to ordered universes with a single dyadic ``immediate successor'' predicate. He associates a nonnegative matrix $A$ with the graph of each constituent, where $\mathbf{e}^\top A^k \mathbf{e}$ yields the number of distinct paths of length $k$. For consistent constituents, these matrices are primitive and irreducible with a unique dominant eigenvalue $\omega_j$. As the universe size $N \to \infty$:
\[
b(e \mid C_j) \approx \frac{1}{\omega_j^n}, \qquad b(C_j \mid e) \approx \frac{1/\omega_j^n}{\sum_i 1/\omega_i^n}.
\]
This is independent of $N$ and depends only on the evidence size $n$ and the constituent's dominant eigenvalue. Constituents with smaller $\omega_j$ (fewer transition possibilities, hence simpler) dominate asymptotically.

\subsubsection{Model~4: Indicator Likelihood (Direct Uniformity)}
\label{sec:indicator}

The simplest model treats evidence as exact logical evidence:
\[
p(e \mid C^w) := \mathbf{1}\{C^w \models e\}.
\]
Combined with the uniform prior $p(C^w) = 1/2^K$:
\[
p(C^w \mid e) = \frac{\mathbf{1}\{C^w \models e\}}{|\mathcal{C}(e)|}, p(e) = \frac{|\mathcal{C}(e)|}{2^K}, p(h \mid e) = \frac{|\mathcal{C}(e \wedge h)|}{|\mathcal{C}(e)|}.
\]
This distinguishes only between evidence that is possible and evidence that is impossible under a world, without distinguishing between evidence that is merely compatible and evidence that is actually likely. It is the most tractable model and is useful when all predicate slots are fully observed, but it is too crude when a statistically principled ranking of compatible constituents is needed. As all other models, in particular Model 2, yields untractable optimization objectives that are combinatorially prohibitive, Model 4 is adopted in the present paper.

\subsection{Handling Partial Evidence: Latent-Variable Models and EM}

When evidence is partial---i.e., the observed predicates do not determine a unique Q-sentence---the underlying Q-sentence is treated as a latent variable. This section details the latent-variable framework and its connections to expectation-maximisation (EM), mixture models, and Markov chain conditional independence.

\subsubsection{The Latent-Q Model as a Finite Mixture}

For each observed evidence item $e_t$, introduce a latent variable $Z_t \in \mathcal{Q}$ representing the underlying complete Q-sentence. The model has the standard finite-mixture architecture:
\[
p(e_t \mid C^w) = \sum_{i=1}^{|\mathcal{Q}|} \underbrace{p(Z_t = i \mid C^w)}_{\text{mixing weight}} \;\underbrace{p(e_t \mid Z_t = i, C^w)}_{\text{emission probability}}.
\]
Here $p(Z_t = i \mid C^w) = \theta_i(C^w)$ is the probability that the latent Q-sentence is $Q_i$ under constituent $C^w$, and $p(e_t \mid Z_t = i, C^w)$ is the \emph{emission probability}---the probability of observing the partial evidence $e_t$ given that the complete underlying pattern is $Q_i$. Under a compatibility model, the emission is deterministic: $p(e_t \mid Q_i) = \mathbf{1}\{Q_i \text{ compatible with } e_t\}$. The hidden Q-sentence $Z_t$ plays exactly the role of the latent component label in a standard mixture model; the difference is interpretational---the hidden class is not a Gaussian component, but a complete logical relation pattern.

For multiple observations $e_1, \dots, e_T$ assumed conditionally independent given $C^w$, the marginal likelihood is:
\[
p(e_{1:T} \mid C^w) = \prod_{t=1}^{T} \sum_{i \in \mathcal{Q}(e_t)} \theta_i(C^w),
\]
where $\mathcal{Q}(e_t) = \{Q_i : Q_i \text{ compatible with } e_t\}$.

\subsubsection{Posterior Responsibilities}

The \emph{posterior responsibility} of Q-sentence $Q_i$ for explaining observation $e_t$ under constituent $C^w$ is:
\[
\gamma_{t,i}^{(w)} := p(Z_t = i \mid e_t, C^w) = \frac{p(e_t \mid Q_i)\,\theta_i(C^w)}{\sum_{j} p(e_t \mid Q_j)\,\theta_j(C^w)}.
\]
This is directly analogous to the responsibilities $\gamma_{ti} = p(z_t = i \mid x_t, \theta)$ in a Gaussian mixture model. It measures how much ``credit'' each hidden Q-sentence deserves for the observed partial evidence. Under the compatibility emission model with uniform mixing weights over the $m_w$ allowed Q-sentences, this reduces to:
\[
\gamma_{t,i}^{(w)} = \frac{\mathbf{1}\{Q_i \in \mathcal{Q}(e_t)\}}{|\mathcal{Q}(e_t) \cap \mathrm{supp}(\theta^{(w)})|},
\]
which numerically coincides with equal splitting when all compatible Q-sentences are equally weighted. But conceptually, this is a \emph{posterior expectation} of an unobserved complete-data assignment, not an arbitrary fractional allocation.

\emph{Expected Q-counts} replace hard counts:
\[
\hat{N}_i^{(w)} = \sum_{t=1}^{T} \gamma_{t,i}^{(w)} = \sum_{t=1}^{T} p(Z_t = i \mid e_t, C^w).
\]
If the latent assignment $Z_t$ were observed, the complete-data count would be $N_i = \sum_t \mathbf{1}\{Z_t = i\}$; the expected count $\hat{N}_i^{(w)}$ is its conditional expectation given the observed partial evidence.

\subsubsection{Expectation-Maximisation (EM) Procedure}

When the constituent-conditioned Q-distributions $\theta^{(w)} = (\theta_1^{(w)}, \dots, \theta_{|\mathcal{Q}|}^{(w)})$ are known, one can compute the posterior over constituents directly. However, when these distributions are unknown and must be estimated from data, the framework becomes a genuine EM algorithm:

\textbf{E-step.} For each observation $e_t$ and each constituent $C^w$, compute the posterior responsibilities:
\[
\gamma_{t,i}^{(w)} = p(Z_t = i \mid e_t, C^w, \theta^{(w)}).
\]
Then compute expected Q-counts: $\hat{N}_i^{(w)} = \sum_t \gamma_{t,i}^{(w)}$.

\textbf{M-step.} Update the mixing weights (and possibly emission parameters) to maximise the expected complete-data log-likelihood:
\[
\theta_i^{(w)} \leftarrow \frac{\hat{N}_i^{(w)}}{\sum_j \hat{N}_j^{(w)}}.
\]
If a Dirichlet prior $\mathrm{Dir}(\alpha_1, \dots, \alpha_{|\mathcal{Q}|})$ is placed on $\theta^{(w)}$, the M-step becomes $\theta_i^{(w)} \propto \hat{N}_i^{(w)} + \alpha_i - 1$ (MAP estimate).

\textbf{Iteration.} The E-step and M-step are alternated until convergence. The incomplete-data log-likelihood $\log p(e_{1:T} \mid C^w)$ is guaranteed to be non-decreasing at each iteration.

The ELBO identity underlying EM is: for any distribution $q(z)$ over the latent assignments,
\[
\log p(e \mid C) = \underbrace{\mathbb{E}_q[\log p(e, z \mid C)] + H(q)}_{\text{ELBO}} \;+\; D_{\mathrm{KL}}(q \| p(z \mid e, C)).
\]
The E-step sets $q(z) = p(z \mid e, C)$, making the KL term zero and the ELBO tight. Any other choice of $q$ yields a lower bound.

\subsubsection{Variational Approximation with Equal Splitting}
\label{sec:elbo}

When the full E-step posterior is intractable, one adopts a mean-field variational family $q(z) = \prod_\ell q_\ell(z_\ell)$ and chooses a convenient $q$. The \emph{equal-splitting} heuristic sets:
\[
q_\ell(i) = q(z_\ell = i \mid r_\ell = r) := \frac{M_{ri}}{\mathrm{deg}(r)}, \qquad \mathrm{deg}(r) = \sum_{j=1}^{K} M_{rj},
\]
where $M_{ri} \in \{0,1\}$ is the compatibility indicator (1 if Q-type $r$ is compatible with attributive type $i$, 0 otherwise). This is a uniform distribution over compatible attributive types---a surrogate E-step that avoids computing the true posterior.

Under this factorisation, the ELBO decomposes into count form:
\begin{align}
\nonumber \mathrm{ELBO}(q; C) &= \sum_{r=1}^{S} t_r \sum_{i=1}^{K} q(i \mid r)\log\theta_i(C) + \\
\nonumber &\sum_{r=1}^{S} t_r \sum_{i=1}^{K} q(i \mid r)\log\phi_{r|i,C} + \sum_{r=1}^{S} t_r H_r,
\end{align}
where $t_r$ are Q-type counts, $\phi_{r|i,C}$ is the emission probability (zero for incompatible pairs), and $H_r = -\sum_i q(i \mid r)\log q(i \mid r)$ is the per-Q-type entropy correction measuring the ambiguity of the mapping $Q_r \to \{A_i\}$ under equal splitting. The replacement $p(e \mid C) \approx p_{\mathrm{sur}}(\hat{u} \mid C)$ is a \emph{model substitution} (surrogate likelihood), not a true ELBO bound, unless one constructs a consistent generative model making the induced counts sufficient.

\subsubsection{The Hierarchical Latent Model}

A richer model introduces latent attributive constituents as an explicit layer:
\[
C^w \;\longrightarrow\; \mathrm{Ct}_j \;\longrightarrow\; Q_i \;\longrightarrow\; e.
\]
The likelihood becomes:
\[
p(e \mid C^w) = \sum_j p(\mathrm{Ct}_j \mid C^w) \sum_i p(Q_i \mid \mathrm{Ct}_j)\,p(e \mid Q_i),
\]
with joint posterior responsibilities over both latent layers:
\[
p(Q_i, \mathrm{Ct}_j \mid e, C^w) \propto p(e \mid Q_i)\,p(Q_i \mid \mathrm{Ct}_j)\,p(\mathrm{Ct}_j \mid C^w).
\]
The Markov chain conditional independence (Section~\ref{sec:markov}) enables the decomposition $p(C^w \mid e) = \sum_{i,j} p(Q_i \mid e)\,p(\mathrm{Ct}_j \mid Q_i)\,p(C^w \mid \mathrm{Ct}_j)$, but this requires the full generative chain to be specified. The exact likelihood requires summing over all $K^n$ possible type assignments (for $n$ individuals and $K$ attributive types), making it combinatorially expensive.

Two important differences between the simpler Q-type count model and the full hierarchical model should be noted. The simpler model treats observed pair types as conditionally independent draws from a common distribution, ignoring the fact that if the same individual $a_i$ appears in multiple pairs, those pairs are statistically coupled through the shared latent type $T_i$ of $a_i$. The hierarchical model captures this dependence automatically because all pairs involving $a_i$ depend on the same latent variable $T_i$.

\subsection{The Objective and Tractable Surrogates}

The goal-oriented semantic communication objective is $F(e) = p(e)\,p(h \mid e)\,(1 - p(h \mid e))$. Under the combined system ($\alpha = 0$, $\lambda = w$), this factorises as:
\[
F(e) = \left(\prod_{j=1}^{c} \Gamma(n_j + 1)\right) N_h \left(1 - \frac{N_h}{D(c)}\right),
\]
where $D(c) = \sum_{i=0}^{K-c} \binom{K-c}{i} \frac{(c+i-1)!}{(n+c+i-1)!}$ and $N_h = \sum_{w \in I_h} \frac{(w-1)!}{(n+w-1)!}$.

Two independent effects control $F(e)$:

\emph{Weight flatness.} The term $\prod_j \Gamma(n_j + 1)$ is minimised when the induced constituent weights $(n_j)$ are as uniform as possible. In the sparse-mass regime, $\sum_j \log\Gamma(n_j + 1) \approx \text{const} + \frac{\pi^2}{12}\sum_j n_j^2$, so minimising the Gamma product is approximately equivalent to minimising the constituent-level second moment $\sum_j n_j^2$.

\emph{Support size.} Since $D(c)$ is decreasing in $c$, the factor $1 - N_h/D(c)$ is also decreasing in $c$. Hence $F(e)$ decreases with larger support $c$.

\subsubsection{Reduction to a Q-Level Tractable Score}

The constituent-level second moment reduces exactly to a Q-level second moment via the enumeration structure:
\[
\textstyle\sum_C n(C)^2 = 2^{-M}\!\left[\left(\sum_q s(q)\right)^{\!2} + \sum_q s(q)^2\right],
\]
where $s(q)$ is the Q-level weight and $M = |\mathcal{Q}|$. Under the entity-block model with block weights $w_u$, distinct-predicate counts $d_u$, and pairwise overlaps $o_{uv} = |A_u \cap A_v|$:
\[
R(e) = \sum_u w_u^2\,2^{d_u} + 2\sum_{u < v} w_u\,w_v\,2^{|A_u \cap A_v|}.
\]
Smaller $R(e)$ indicates flatter induced constituent weights, providing a feasible ranking criterion without constructing the doubly-exponential constituent space.

\subsubsection{Indicator-Likelihood Surrogate}

Under Model~4, each per-hypothesis term reduces to $\mathcal{F}_i \approx 2^{-\gamma_i}$ where $\gamma_i = 2^{|\mathcal{Q}|-K} - 2^{|\mathcal{Q}|-K-H_i}$. The sum is dominated by $\gamma_{\min}$, and optimisation reduces to a lexicographic comparison key over small integers $(c, K, -H_{(1)}, \dots)$. This method is adopted in the present paper.

\subsection{Summary}

All models share the same logical language structure ($\mathcal{P} \to \mathcal{Q} \to C_j \to C^w$) and the principle that constituent probabilities determine all other sentence probabilities. The hierarchical Bayesian interpretation---with its connections to finite mixture models, EM, latent class models, and constrained multinomials with structural zeros---is the most developed reading, but the logical structure itself does not mandate it. The indicator, eigenvalue, and variational models demonstrate that other probabilistic characterisations of the same logical scaffolding are both possible and useful.

The central design principle across all models is: \emph{logic specifies which configurations are possible; statistics specifies how probable the possible configurations are.} The choice among models is a modelling decision, not a logical necessity.

\end{document}